\documentstyle[sprocl,overcite,epsf]{article}

\def\Preprint{\vspace*{-7.0cm} %\noindent hep-ph/9710XXX 
  \hfill FTUV/97-45 \\ \mbox{}\hfill 
  IFIC/97-46 \\  \mbox{}\hfill November 1997 \\ 
  \vspace{4cm}}
\def\refjl#1#2#3#4#5#6{\bibitem{#1} #2, {\it #3} {\bf #4} (#5) #6.}

\def\etal{{\it et al}}
\def\NP{Nucl. Phys.}
\def\NPPS{Nucl. Phys. B (Proc. Suppl.)}
\def\PL{Phys. Lett.}
\def\PRL{Phys. Rev. Lett.}
\def\PR{Phys. Rev.}
\def\PRep{Phys. Rep.}
\def\ZP{Z. Phys.}

\def\JPG{J. Phys. G: Nucl. Phys.}           %1975--1988
\def\RPP{Rep. Prog. Phys.}

\newcommand{\eqn}[1]{(\ref{#1})}
\newcommand{\be}{\begin{equation}}
\newcommand{\ee}{\end{equation}}
\newcommand{\no}{\nonumber}
\newcommand{\bel}[1]{\be\label{#1}}
\newcommand{\ba}{\begin{array}{c}}
\newcommand{\bat}{\begin{array}{cc}}
\newcommand{\ea}{\end{array}}
\newcommand{\beqn}{\begin{eqnarray}}
\newcommand{\eeqn}{\end{eqnarray}}

\newcommand{\bi}{\begin{itemize}}
\newcommand{\ei}{\end{itemize}}

\newcommand{\rms}{\rm\scriptsize}

\newcommand{\toLow}{\stackrel{q^2 \ll M_W^2}{\,\longrightarrow\,}}

\newcommand{\cL}{{\cal L}}

\newcommand{\cP}{{\cal P}}

\newcommand{\cH}{{\cal H}}
\newcommand{\cA}{{\cal A}}

\begin{document}
\title{PRECISION TESTS OF THE STANDARD MODEL
  \footnote{Lectures given at the 25$^{\rms th}$ Winter Meeting on
  Fundamental Physics (Formigal, 3--8 March 1997)}
}

\author{A. PICH}

\address{Departament de F\'{\i}sica Te\`orica, 
         IFIC,  CSIC --- Universitat de Val\`encia, \\ 
         Dr. Moliner 50, E--46100 Burjassot, Val\`encia, Spain}

\maketitle
%\Preprint
\setcounter{footnote}{0}

\abstracts{
Precision measurements of electroweak observables provide
stringent tests of the Standard Model structure and an
accurate determination of its parameters.
An overview of the present experimental status is presented. 
}

\Preprint
  
\section{Introduction}
\label{sec:introduction}

The Standard Model (SM) constitutes one of the most successful achievements
in modern physics. It provides a very elegant theoretical
framework, which is able to describe all known experimental
facts in particle physics.
A detailed description of the SM
and its impressive phenomenological success
can be found in Refs.~\citen{jaca:94}
and \citen{sorrento:94}, which discuss the electroweak and strong
sectors, respectively.

The high accuracy achieved by the most recent experiments
allows to make stringent tests of the SM structure at the level
of quantum corrections.
The different measurements complement each other in their different
sensitivity to the SM parameters. 
Confronting these measurements with the theoretical predictions, one
can check the internal consistency of the SM framework and
determine its parameters.

These lectures provide an overview of our
present experimental knowledge on the electroweak couplings.
A brief description of some classical QED tests is presented
in Section~\ref{sec:qed}.
The leptonic couplings of the $W^\pm$ bosons are analyzed in
Section~\ref{sec:cc}, where the tests on lepton universality
and the Lorentz structure of the $l^-\to\nu_l l'^-\bar\nu_{l'}$
transition amplitudes are discussed.
Section~\ref{sec:nc} describes the status of the
neutral--current sector, using the latest experimental results reported by
LEP and SLD.
Some summarizing comments are finally given in Section~\ref{sec:summary}.
I have skipped completely the analysis of the $W^\pm$ couplings
to the charged quark currents; a rather exhaustive description
of the existing constraints on the quark--mixing
matrix and the present status of CP--violation phenomena has been
given in Refs.~\citen{comillas:95} and \citen{win:97}.

\section{QED}
\label{sec:qed}

A general description of the electromagnetic coupling of a 
spin--$\frac{1}{2}$ charged lepton to the virtual photon
involves three different form factors:
\bel{eq:em_ff}
T[l\bar l \gamma^*] = e \, \varepsilon_\mu(q) \, \bar l
\left[F_1(q^2)\gamma^\mu
+ i{F_2(q^2)\over 2 m_l} \sigma^{\mu\nu}q_\nu +
{F_3(q^2)\over 2 m_l} \sigma^{\mu\nu}\gamma_5 q_\nu\right] l \ ,
\ee
where $q^\mu$ is the photon momentum.
Owing to the conservation of the electric charge,
$F_1(0)=1$.
At $q^2=0$, the other two form factors reduce to the
lepton 
magnetic dipole moment 
$\mu_l\equiv (e /2 m_l) \, (g_l/2) = e (1+F_2(0))/2 m_l$,
and electric dipole moment $d_l = e F_3(0)/2 m_l$.

The $F_i(q^2)$ form factors are sensitive quantities to a
possible lepton substructure.
Moreover, $F_3(q^2)$ violates $T$ and $P$ invariance; 
thus, the electric dipole moments, which vanish in the SM, 
constitute a good probe of CP violation.
Owing to their chiral--changing structure, the 
magnetic and electric dipole moments 
may provide important insights on the mechanism responsible for
mass generation. In general, one expects \cite{MA:94}
that a fermion of mass
$m_f$ (generated by physics at some scale $M\gg m_f$) will have
induced dipole moments proportional to some power of $m_f/M$.
%Therefore, the $\tau$ should be a good testing ground for this kind of effects.

The measurement of the $e^+e^-\to l^+ l^-$ cross-section has been
used to test the universality of the leptonic QED couplings.
At low energies, where the $Z$ contribution is small, the deviations
from the QED prediction are usually parameterized through\footnote{
A slightly different parameterization is adopted 
for $e^+e^-\to e^+e^-$, to account for
the $t$--channel contribution \protect\cite{MA:90}.}
\be
\sigma(e^+e^-\to l^+ l^-)\, =\, \sigma_{\mbox{\rms QED}} \,
\left( 1 \mp {s\over s-\Lambda_\pm^2}\right)^2.
\ee
The cut-off parameters $\Lambda_\pm$ characterize the validity of QED
and measure the point-like nature of the leptons.
From PEP and PETRA data, one finds \cite{MA:90}:
$\Lambda_+(e)> 435$ GeV, $\Lambda_-(e)> 590$ GeV,
$\Lambda_+(\mu)> 355$ GeV,  $\Lambda_-(\mu)> 265$ GeV,
$\Lambda_+(\tau)> 285$ GeV and 
$\Lambda_-(\tau)> 246$ GeV (95\% CL),
which correspond to upper limits on the lepton charge radii
of about $10^{-3}$ fm.

%%%%%%%%%%%%% FIGURE %%%%%%%%%%%%%%%%%
\begin{figure}[bth]
\centering
\vspace{0.3cm}
%\framebox(50,20){FIGURE}
\centerline{\epsfxsize =\linewidth \epsfbox{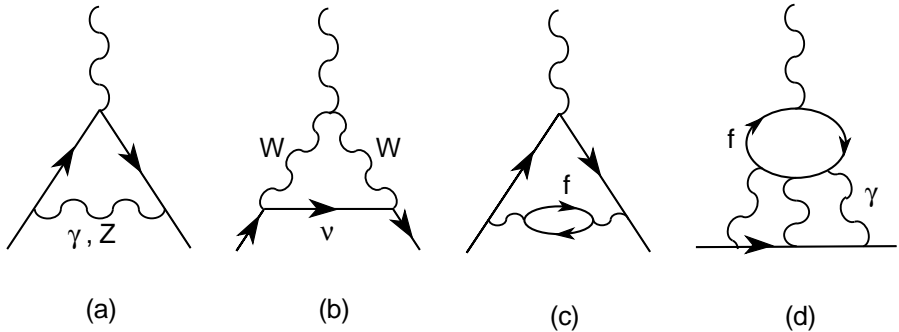}}
%\vspace{-0.5cm}
\caption{Some Feynman diagrams contributing to $a_l$.
\label{fig:AnMagMom}}
\end{figure}
%%%%%%%%%%%%%%%%%%%%%%%%%%%%%%%%%%%%%%%%%

The most stringent QED test comes from the high--precision
measurements of the $e$ and $\mu$ anomalous magnetic moments 
\cite{KI:90,KI:96,CKM:96,PPR:95,KKSS:92,dRA:94,HK:97,HKS:96,KS:95,BPP:96,KR:97,ADH:97}
$a_l\equiv (g_l-2)/2$:
\beqn\label{eq:a_e}
a_e&=&\left\{ \bat
(115 \, 965 \, 214.0\pm 2.8) \times 10^{-11} & (\mbox{\rm Theory})
\\
(115 \, 965 \, 219.3\pm 1.0) \times 10^{-11} & (\mbox{\rm Experiment})
\ea \, , \right.\\ 
a_\mu&=&\left\{ \bat
%(1 \, 165 \, 919.2\pm 1.9) 
(1 \, 165 \, 917.1\pm 1.0)   
\times 10^{-9} & (\mbox{\rm Theory})
\\
(1 \, 165 \, 923.0\pm 8.4) \times 10^{-9} & (\mbox{\rm Experiment})
\ea \, . \right.  
\eeqn
The impressive agreement between theory and experiment 
(at the level of the ninth digit for $a_e$)
promotes QED
to the level of the best theory ever build by the human mind
to describe nature.
Hypothetical {\it new--physics} effects are 
constrained to the ranges
$|\delta a_e | < 1.1 \times 10^{-10}$ and
$|\delta a_\mu | < 2.2 \times 10^{-8}$ (95\% CL).

To a measurable level, $a_e$ arises entirely from virtual electrons and
photons; these contributions are known \cite{KI:96} to $O(\alpha^4)$. 
The sum of all other QED corrections, associated with higher--mass
leptons or intermediate quarks, only amounts to 
$+(0.4366\pm 0.0042)\times 10^{-11}$,
while the weak interaction effect is a tiny $+0.0030\times 10^{-11}$;
these numbers \cite{KI:96} are well below the present 
experimental precision.
The theoretical error is dominated by the uncertainty in the
input value of the electromagnetic coupling $\alpha$. In fact, 
turning things around, one can use $a_e$ to make the most precise
determination of the fine structure constant \cite{KI:96}:
\bel{eq:alpha}
\alpha^{-1} = 137.03599993 \pm 0.00000052 \, .
\ee
The resulting accuracy is one order of magnitude better than
the usually quoted value \cite{PDG:96}
$\alpha^{-1} = 137.0359895 \pm 0.0000061$.

The anomalous magnetic moment of the muon is sensitive to virtual 
contributions from heavier states; compared to $a_e$, they scale
as $m_\mu^2/m_e^2$.
The main theoretical uncertainty on $a_\mu$ has a QCD origin.
Since quarks have electric charge, virtual quark--antiquark pairs
can be created by the photon leading to the so--called
{\it hadronic vacuum polarization} corrections to the photon propagator
(Figure 1.c).
Owing to the non-perturbative character of QCD at low energies,
the light--quark contribution cannot be reliably calculated at
present; fortunately, this effect can be
extracted from the measurement of the cross-section
$\sigma(e^+e^-\to \mbox{\rm hadrons})$  at low energies,
and from the invariant--mass distribution of the final hadrons in
$\tau$ decays \cite{ADH:97}.
The large uncertainties of the present data are the dominant
limitation to the achievable theoretical precision on $a_\mu$.
It is expected that this will be improved at the DA$\Phi$NE
$\Phi$ factory, where an accurate measurement of the hadronic production
cross-section in the most relevant kinematical region is expected
\cite{daphne}.
Additional QCD uncertainties stem from the (smaller)
{\it light--by--light scattering}
contributions, where four photons couple to a light--quark loop
(Figure 1.d);
these corrections are under active investigation at present
\cite{dRA:94,HK:97,HKS:96,KS:95,BPP:96}.

The improvement of the theoretical $a_\mu$ prediction is of great interest
in view of the new E821 experiment \cite{BNL:E821}, presently running
at Brookhaven,
which aims to reach a sensitivity of at least $4\times 10^{-10}$, 
and thereby observe the contributions from virtual $W^\pm$ and $Z$ bosons
\cite{CKM:96,PPR:95,KKSS:92}
($\delta a_\mu|_{\mbox{\rms weak}} \sim 15 \times 10^{-10}$).
The extent to which this measurement could provide a meaningful
test of the electroweak theory depends critically on
the accuracy one will be able to achieve pinning down the QCD corrections.

Experimentally, very little is known about $a_\tau$ since the spin
precession method used for the lighter leptons cannot be applied
due to the very short lifetime of the $\tau$.
The effect is however visible in the $e^+e^-\to\tau^+\tau^-$
cross-section. The limit $|a_\tau|<0.023$ (95\% CL) has been derived
\cite{SI:83}  %,MA:89}
from PEP and PETRA data. This limit actually probes the corresponding
form factor $F_2(s)$ at $s\sim 35$ GeV.
A more direct bound at $q^2=0$ has been extracted \cite{tau96:L3}
from the decay $Z\to\tau^+\tau^-\gamma$:
\be
|a_\tau| < 0.049   \qquad (90\%\,\mbox{\rm CL}) \, .
%   0.0104 \qquad (95\%\,\mbox{\rm CL}) \, .
\ee
A better, but more model--dependent, limit has been obtained
\cite{EM:93} from the $Z\to\tau^+\tau^-$ decay width:
$-0.004 < a_\tau < 0.006$.

In the SM the overall value of $a_\tau$ is dominated by the second order
QED contribution \cite{SC:48},
$a_\tau \approx \alpha / 2 \pi$.
Including QED corrections up to O($\alpha^3$),
hadronic vacuum polarization contributions
and the corrections due to the weak interactions 
(which are a factor 380
larger than for the muon), the tau anomalous magnetic moment has been
estimated to be \cite{NA:78,SLM:91}
\bel{eq:a_th_tau}
a_\tau\big |_{\mbox{\rms th}} \, = \, (1.1773 \pm 0.0003)
     \times 10^{-3} \, .
\ee

So far, no evidence has been found for any CP--violation signature
in the lepton sector.
The present limits on the leptonic electric dipole moments are
\cite{PDG:96,tau96:L3}:
$$
d_e \, = \,  (-0.27\pm 0.83)\times 10^{-26}\, e\,\mbox{\rm cm},
$$
\be
d_\mu \, = \,  (3.7\pm 3.4)\times 10^{-19}\, e\,\mbox{\rm cm},
\ee
$$
|d_\tau| \, < \,  2.7\times 10^{-16}\, e\,\mbox{\rm cm}
   \qquad (90\%\,\mbox{\rm CL}) \, .
   %5.8\times 10^{-17}\, e\,\mbox{\rm cm}.
$$

\section{Charged--Current Couplings}
\label{sec:cc}

In the SM, the charged--current interactions are governed by an
universal coupling $g$:
\bel{eq:cc_mixing}
\cL_{\mbox{\rms CC}} = {g\over 2\sqrt{2}}\,
W^\dagger_\mu\,\left[\sum_{ij}\,
\bar u_i\gamma^\mu(1-\gamma_5) V_{ij} d_j 
\, +\,\sum_l\, \bar\nu_l\gamma^\mu(1-\gamma_5) l
\right]\, + \, \mbox{\rm h.c.}\; .
\ee
In the original basis of weak eigenstates quarks and leptons have
identical interactions. The diagonalization of the fermion masses
gives rise to the unitary quark mixing matrix $V_{ij}$, which couples
any {\it up--type} quark with all {\it down--type} quarks. 
For massless neutrinos, the analogous leptonic mixing matrix can
be eliminated by a redefinition of the neutrino fields.
The lepton flavour is then conserved in the minimal SM without
right--handed neutrinos.

\subsection{$\mu^-\to e^-\bar\nu_e\nu_\mu$}
\label{subsec:MuDecay}

%%%%%%%%%%%%%%%%   FIGURES 1 and 2 %%%%%%%%%%%%%%%%%%%
\begin{figure}[bth]
\vfill
\centerline{
\begin{minipage}[t]{.4\linewidth}\centering
%\framebox(50,20){FIGURE}
{\epsfysize =2.7cm \epsfbox{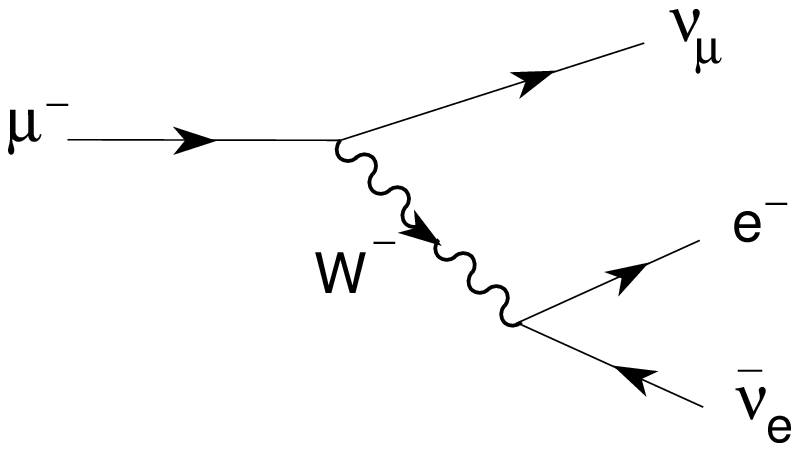}}
\caption{$\mu$--decay diagram.}
\label{fig:mu_decay}
\end{minipage}
\hspace{0.6cm}
\begin{minipage}[t]{.54\linewidth}\centering
%\framebox(50,20){FIGURE}
%\epsfig{file=TauDecay2.ps,height=35mm}
{\epsfysize =2.7cm \epsfbox{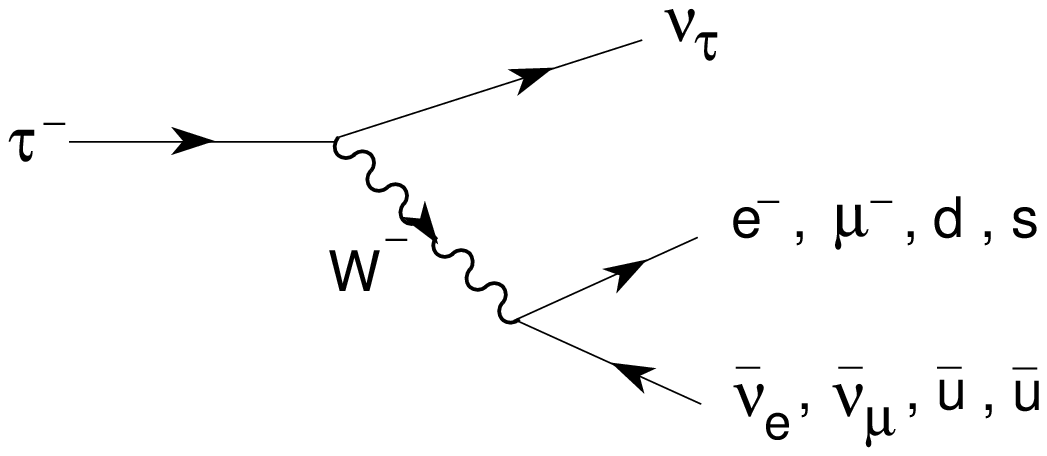}}
\caption{$\tau$--decay diagram.}
\end{minipage}
}
\vfill
\end{figure}
%%%%%%%%%%%%%%%%%%%%% END FIGURES %%%%%%%%%%%%%%%%%%%%%%%%%%

The simplest flavour--changing process is the leptonic
decay of the muon, which proceeds through the $W$--exchange
diagram shown in Figure~\ref{fig:mu_decay}.
The momentum transfer carried by the intermediate $W$ is very small
compared to $M_W$. Therefore, the vector--boson propagator reduces
to a contact interaction,
\bel{eq:low_energy}
{-g_{\mu\nu} + q_\mu q_\nu/M_W^2 \over q^2-M_W^2}\quad\;
 \toLow\quad\; {g_{\mu\nu}\over M_W^2}\, .
\ee
The decay can then be described through an effective local
4--fermion Hamiltonian,
\bel{eq:mu_v_a}
\cH_{\mbox{\rms eff}}\, = \, {G_F \over\sqrt{2}}
\left[\bar e\gamma^\alpha (1-\gamma_5) \nu_e\right]\,
\left[ \bar\nu_\mu\gamma_\alpha (1-\gamma_5)\mu\right]\, , 
\ee
where
\bel{eq:G_F}
{G_F\over\sqrt{2}} = {g^2\over 8 M_W^2}
\ee
is called the Fermi coupling constant.
$G_F$ 
is fixed by the total decay width,
\bel{eq:mu_lifetime}
{1\over\tau_\mu}\, = \, \Gamma(\mu^-\to e^-\bar\nu_e\nu_\mu)
\, = \, {G_F^2 m_\mu^5\over 192 \pi^3}\,
\left( 1 + \delta_{\mbox{\rms RC}}\right) \, 
f\left(m_e^2/m_\mu^2\right) \, ,
\ee
where
$\, f(x) = 1-8x+8x^3-x^4-12x^2\ln{x}$,
and
\bel{eq:qed_corr}
(1+\delta_{\mbox{\rms RC}})  =  
\left[1+{\alpha(m_\mu)\over 2\pi}\left({25\over 4}-\pi^2\right)\right]\,
\left[ 1 +{3\over 5}{m_\mu^2\over M_W^2} - 2 {m_e^2\over M_W^2}\right]
=  0.9958 \, 
\ee
takes into account the leading higher--order corrections \cite{KS:59,MS:88}.
The measured lifetime \cite{PDG:96},
$\tau_\mu=(2.19703\pm 0.00004)\times 10^{-6}$ s,
implies the value
\bel{eq:gf}
G_F\, = \, (1.16639\pm 0.00002)\times 10^{-5} \:\mbox{\rm GeV}^{-2}
\,\approx\, {1\over (293 \:\mbox{\rm GeV})^2} \, .
\ee

\subsection{$\tau$ Decay}
\label{subsec:TauDecay}

The decays of the $\tau$ proceed through the same
$W$--exchange mechanism as the leptonic $\mu$ decay.
The only difference is that several final states
are kinematically allowed:
$\tau^-\to\nu_\tau e^-\bar\nu_e$,
$\tau^-\to\nu_\tau\mu^-\bar\nu_\mu$,
$\tau^-\to\nu_\tau d\bar u$ and $\tau^-\to\nu_\tau s\bar u$.
Owing to the universality of the $W$--couplings, all these
decay modes have equal amplitudes (if final fermion masses and
QCD interactions are neglected), except for an additional
$N_C |V_{ui}|^2$ factor ($i=d,s$) in the semileptonic
channels, where $N_C=3$ is the number of quark colours. 
Making trivial kinematical changes in Eq.~\eqn{eq:mu_lifetime},
one easily gets the lowest--order prediction for the total
$\tau$ decay width:
\bel{eq:tau_decay_width}
{1\over\tau_\tau}\equiv\Gamma(\tau) \approx
\Gamma(\mu) \left({m_\tau\over m_\mu}\right)^5
\left\{ 2 + N_C 
\left( |V_{ud}|^2 + |V_{us}|^2\right)\right\}
\approx {5\over\tau_\mu}\left({m_\tau\over m_\mu}\right)^5 .
\ee
From the measured muon lifetime, one has then
$\tau_\tau\approx 3.3\times 10^{-13}$ s, to be compared
with the experimental value \cite{PDG:96,ALEPH:97}
$\tau_\tau^{\mbox{\rms exp}} = (2.900\pm 0.012)\times 10^{-13}$ s.

%%%%%%%%%%%%%%% Table Tau Properties %%%%%%%%%%%%%%%%%%%%%%%%%%%%%%
%
\begin{table}[thb]
\centering
\caption{Experimental values \protect\cite{PDG:96}
of some basic $\tau$ decay branching fractions.}
%$h^-$ stands for either $\pi^-$ or $K^-$.}
\label{tab:parameters}
\vspace{0.2cm}
\begin{tabular}{|c|c|}
\hline 
%$m_\tau$ & $(1777.00^{+0.30}_{-0.27})$ MeV \\
%$\tau_\tau$ & $(290.21\pm 1.15)$ fs \\
$B_{\tau\to e}$ & $(17.80\pm 0.08)\% $ \\
$B_{\tau\to\mu}$ & $(17.30\pm 0.10)\% $ \\
$R_\tau^B \equiv (1-B_{\tau\to e}-B_{\tau\to\mu})/ B_{\tau\to e}$ &
 $3.646\pm 0.022$ \\
Br($\tau^-\to\nu_\tau\pi^-$) & $(11.07\pm 0.18)\% $ \\
Br($\tau^-\to\nu_\tau K^-$) & $(0.71\pm 0.05)\% $ \\
%Br($\tau^-\to\nu_\tau h^-$) & $(11.70\pm 0.11)\% $
\hline
\end{tabular}
\end{table}
%
%%%%%%%%%%%%%%%%%%%%%%%%%%%%%%%%%%%%%%%%%%%%%%%%%%%%%%%%%%%%%%%%%%%%%

The branching ratios into the different decay modes are
predicted to be:
$$
B_{\tau \to l} \equiv
\mbox{\rm Br}(\tau^-\to\nu_\tau l^-\bar\nu_l) \approx {1\over 5} = 20\%\ , 
$$
\bel{eq:tau_br}
 R_\tau\equiv {\Gamma(\tau\to\nu_\tau + \mbox{\rm hadrons})\over 
\Gamma(\tau^-\to\nu_\tau e^-\bar\nu_e)} \approx N_C\, ,
\ee
in good agreement with the measured numbers \cite{PDG:96}, 
given in Table~\ref{tab:parameters}. 
Our naive predictions only deviate
from the experimental results by about 20\%. This
is the expected size of the corrections induced by
the strong interactions between the final quarks, 
that we have neglected.
Notice that the measured $\tau$ hadronic width provides strong evidence
for the colour degree of freedom.

The pure leptonic decays 
$\tau^-\to e^-\bar\nu_e\nu_\tau,\mu^-\bar\nu_\mu\nu_\tau$
are theoretically understood at the level of the electroweak
radiative corrections \cite{MS:88}.
The corresponding decay widths are given by Eqs.~\eqn{eq:mu_lifetime}
and \eqn{eq:qed_corr},
making the appropriate changes for the masses of the initial and final
leptons.

Using the value of $G_F$  
measured in $\mu$ decay, Eq.~\eqn{eq:mu_lifetime} 
provides a relation between the $\tau$ lifetime
and the leptonic branching ratios \cite{tau96}:
\be\label{eq:relation}
B_{\tau\to e} =  {B_{\tau\to \mu} \over 0.972564\pm 0.000010} = 
{ \tau_{\tau} \over (1.6321 \pm 0.0014) \times 10^{-12}\, \mbox{\rm s} }
\, .
\ee
The errors reflect the present uncertainty of $0.3$ MeV
in the value of $m_\tau$.

%%%%%%%%%%%%% FIGURE %%%%%%%%%%%%%%%%%
\begin{figure}[bt] %h]
\centering
\vspace{0.3cm}
%\framebox(50,20){FIGURE}
\centerline{\epsfxsize =9cm \epsfbox{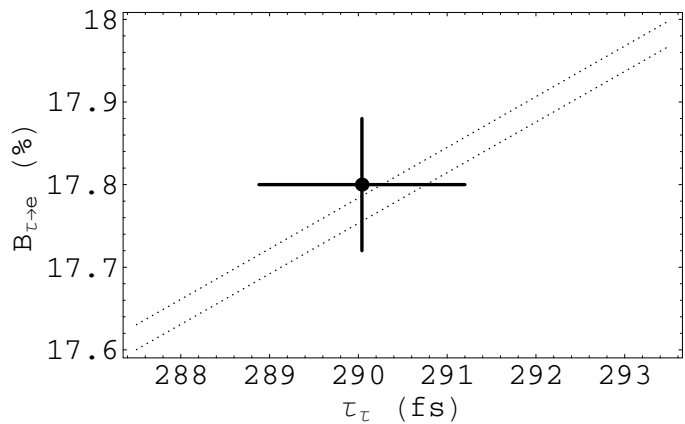}}
%\vspace{-0.5cm}
\caption{Relation between $B_{\tau\to e}$ and $\tau_\tau$. The dotted
band corresponds to Eq.~(\protect\ref{eq:relation}).
\label{fig:BeLife}}
\end{figure}
%%%%%%%%%%%%%%%%%%%%%%%%%%%%%%%%%%%%%%%%%

The predicted $B_{\tau\to\mu}/B_{\tau\to e}$ 
ratio is in perfect agreement with the measured
value $B_{\tau\to\mu}/B_{\tau\to e} = 0.972 \pm 0.007$.  As shown in
Figure~\ref{fig:BeLife}, the relation between $B_{\tau\to e}$ and
$\tau_\tau$ is also well satisfied by the present data. Notice, that this
relation is very sensitive to the value of the $\tau$ mass
[$\Gamma(\tau^-\to l^-\bar\nu_l\nu_\tau)\propto m_\tau^5$]. 
% The most recent measurements of
% $\tau_\tau$, $B_e$ and $m_\tau$ have consistently moved the world averages
% in the correct direction, eliminating the previous ($\sim 2\sigma$)
% disagreement \cite{PI:92}. 
The experimental precision (0.4\%) is already approaching the
level where a possible non-zero $\nu_\tau$ mass could become relevant; the
present bound \cite{tau96:PA}
$m_{\nu_\tau}< 18.2$ MeV (95\% CL) only guarantees that such 
effect is below 0.08\%.

\subsection{Semileptonic Decays}
\label{subsec:Semileptonic}

Semileptonic decays such as $\tau^-\to\nu_\tau P^-$ or 
$P^-\to l^-\bar\nu_l$ [$P=\pi,K$] can be predicted
in a similar way. The effects of the strong interaction
are contained in the so--called decay constants $f_P$,
which parameterize the hadronic matrix element of the
corresponding weak current:
\bel{eq:f_pi_K}\begin{array}{ccc}
\langle\pi^-(p)|\bar d \gamma^\mu\gamma_5 u | 0 \rangle &
\equiv & -i \sqrt{2} f_\pi p^\mu \, , \\
\langle K^-(p)|\bar s \gamma^\mu\gamma_5 u | 0 \rangle &
\equiv & -i \sqrt{2} f_K p^\mu \, . 
\ea\ee

Taking appropriate ratios of different semileptonic decay widths
involving the same meson $P$, the dependence on the decay
constants factors out. Therefore, those ratios can be predicted
rather accurately:
\beqn\label{eq:r_l_P}
R_{\pi\to e/\mu}& \!\!\!\equiv &\!\!\! {\Gamma(\pi^-\to e^-\bar\nu_e)\over
\Gamma(\pi^-\to \mu^-\bar\nu_\mu)}\, = \,
{m_e^2 (1-m_e^2/m_\pi^2)^2\over m_\mu^2 (1-m_\mu^2/m_\pi^2)^2}
\, (1 + \delta R_{\pi\to e/\mu})
\no\\ & \!\!\! = &\!\!\!
(1.2351\pm 0.0005)\times 10^{-4} ,
\no\\
R_{\tau/\pi} & \!\!\!\equiv &\!\!\!
 {\Gamma(\tau^-\to\nu_\tau\pi^-) \over
 \Gamma(\pi^-\to \mu^-\bar\nu_\mu)} \, = \,
%\Big\vert {g_\tau\over g_\mu}\Big\vert^2
{m_\tau^3\over 2 m_\pi m_\mu^2}
{(1-m_\pi^2/ m_\tau^2)^2\over
 (1-m_\mu^2/ m_\pi^2)^2} 
\left( 1 + \delta R_{\tau/\pi}\right) 
\no\\ & \!\!\! = &\!\!\!
9774\pm 15 \, ,
\\ 
R_{\tau/K} &\!\!\! \equiv &\!\!\! {\Gamma(\tau^-\to\nu_\tau K^-) \over
 \Gamma(K^-\to \mu^-\bar\nu_\mu)} \, = \,
%\Big\vert {g_\tau\over g_\mu}\Big\vert^2
{m_\tau^3\over 2 m_K m_\mu^2}
{(1-m_K^2/m_\tau^2)^2\over
(1-m_\mu^2/ m_K^2)^2} 
\left( 1 + \delta R_{\tau/K}\right) 
\no\\ & \!\!\! = &\!\!\!
 480.4\pm 1.1 \, , \no
\eeqn
where $\delta R_{\pi\to e/\mu}= - (3.76\pm 0.04)\% $,
$\delta R_{\tau/\pi} = (0.16\pm 0.14)\% $ and
$\delta R_{\tau/K} = (0.90\pm 0.22)\%  $
are the computed \cite{MS:93,DF:94} radiative corrections.
These predictions are in excellent agreement with the
measured ratios \cite{PDG:96,ALEPH:97,BR:92,CZ:93}:
$R_{\pi\to e/\mu} = (1.2310\pm 0.0037)\times 10^{-4}$, 
$R_{\tau/\pi} = 9937\pm 166$ and
$R_{\tau/K} = 477\pm 34$.

\subsection{Universality Tests}
\label{subsec:Universality}

%%%%%%%%%%%%%%%%  TABLES  %%%%%%%
\begin{table}[tbh]
\centering
\caption{Present constraints on $|g_\mu/g_e|$.}
\label{tab:univme}
\vspace{0.2cm}
\begin{tabular}{|c|c|}
\hline
& $|g_\mu/g_e|$ \\ \hline
$B_{\tau\to\mu}/B_{\tau\to e}$ & $0.9997\pm 0.0037$
\\
$R_{\pi\to e/\mu}$ & $1.0017\pm 0.0015$
\\
$\sigma\cdot B_{W\to\mu/e}$ \ \ ($p\bar p$) & $0.98\pm 0.03$
\\
$B_{W\to\mu/e}$ (LEP2) & $0.92\pm 0.08$
\\ \hline
\end{tabular}\vspace{1cm}
%\end{table}
%
%%%%%%%%%%%%%%
%\begin{table}[bth]
%\centering
\caption{Present constraints on $|g_\tau/g_\mu|$.}
\label{tab:univtm}
\vspace{0.2cm}
\begin{tabular}{|c|c|}
\hline
& $|g_\tau/g_\mu|$  \\ \hline
$B_{\tau\to e}\tau_\mu/\tau_\tau$ & $1.0008\pm 0.0030$
\\
$R_{\tau/\pi}$ &  $1.008\pm 0.008$
\\
$R_{\tau/K}$ & $0.997\pm 0.035$
\\
%$R_{\tau/h}$ & $1.004\pm 0.005$
%\\
$\sigma\cdot B_{W\to\tau/\mu}$ \ \ ($p\bar p$) & $1.02\pm 0.05$
\\
$B_{W\to\tau/\mu}$ (LEP2) & $1.18\pm 0.11$
\\ \hline
\end{tabular}
\end{table}
%
%%%%%%%%%%%%%%%%%%%%%%%%%%%%%%%%%%%%%%%%%%%%%%%%

All these measurements can be used to test the universality of
the $W$ couplings to the leptonic charged currents.
Allowing the coupling $g$ in Eq.~\eqn{eq:cc_mixing}
to depend on the considered lepton flavour 
(i.e.  $g_e$, $g_\mu$, $g_\tau$), 
the ratios $B_{\tau\to\mu}/B_{\tau\to e}$ and $R_{\pi\to e/\mu}$ 
constrain $|g_\mu/g_e|$, while
$B_{\tau\to e}/\tau_\tau$ and $R_{\tau/P}$
provide information on $|g_\tau/g_\mu|$.
The present results are shown in Tables \ref{tab:univme} and
\ref{tab:univtm}, together with the values obtained from 
the comparison of the $\sigma\cdot B$ partial production
cross-sections for the various $W^-\to l^-\bar\nu_l$ decay
modes at the $p\bar p$ colliders \cite{PDG:96},
and from the most recent LEP2 measurements of the leptonic
$W^\pm$ branching ratios \cite{lepewwg:97}.

Although the direct constraints from the measured $W^-\to l^-\bar\nu_l$
branching ratios are meager, the indirect information obtained in
$W^\pm$--mediated decays provides stringent tests of the
$W^\pm$ interactions.
The present data verify the universality of the leptonic
charged--current couplings to the 0.15\% ($\mu/e$) and 0.30\%
($\tau/\mu$) level. The precision of the most recent
$\tau$--decay measurements is becoming competitive with the 
more accurate $\pi$--decay determination. 
It is important to realize the complementarity of the
different universality tests. 
The pure leptonic decay modes probe
the charged--current couplings of a transverse $W$. In contrast,
the decays $\pi/K\to l\bar\nu$ and $\tau\to\nu_\tau\pi/K$ are only
sensitive to the spin--0 piece of the charged current; thus,
they could unveil the presence of possible scalar--exchange
contributions with Yukawa--like couplings proportional to some
power of the charged--lepton mass.
One can easily imagine new physics scenarios which would modify 
differently the two types of leptonic couplings \cite{MA:94}. 
For instance,
in the usual two Higgs doublet model, charged--scalar exchange
generates a correction to the ratio $B_{\tau\to\mu}/B_{\tau\to e}$, but 
$R_{\pi\to e/\mu}$ remains unaffected.
Similarly, lepton mixing between the $\nu_\tau$ and an hypothetical
heavy neutrino would not modify the ratios $B_{\tau\to\mu}/B_{\tau\to e}$
and
$R_{\pi\to e/\mu}$, but would certainly correct the relation between
$B_{\tau\to l}$ and the $\tau$ lifetime.

\subsection{Lorentz Structure}
\label{sec:lorentz}

Let us consider the leptonic
decays $l^-\to\nu_l l'^-\bar\nu_{l'}$, 
where the pair ($l$, $l^\prime $)
may be ($\mu$, $e$), ($\tau$, $e$), or ($\tau$, $\mu$). 
The most general, local, derivative--free, lepton--number conserving, 
four--lepton interaction Hamiltonian, 
consistent with locality and Lorentz invariance
\cite{MI:50,BM:57,KS:57,SCH:83,FGJ:86,FG:93,PS:95},
\be
{\cal H} = 4 \frac{G_{l'l}}{\sqrt{2}}
\sum_{n,\epsilon,\omega}          %^{n = S,V,T}
g^n_{\epsilon\omega}   %g^n_{l'_\epsilon l^{\phantom{'}}_\omega}
\left[ \overline{l'_\epsilon} 
\Gamma^n {(\nu_{l'})}_\sigma \right]\, 
\left[ \overline{({\nu_l})_\lambda} \Gamma_n 
	l_\omega \right]\ ,
\label{eq:hamiltonian}
\ee
contains ten complex coupling constants or, since a common phase is
arbitrary, nineteen independent real parameters
which could be different for each leptonic decay.
The subindices
$\epsilon , \omega , \sigma, \lambda$ label the chiralities (left--handed,
right--handed)  of the  corresponding  fermions, and $n$ the
type of interaction: 
scalar ($I$), vector ($\gamma^\mu$), tensor 
($\sigma^{\mu\nu}/\sqrt{2}$).
For given $n, \epsilon ,
\omega $, the neutrino chiralities $\sigma $ and $\lambda$
are uniquely determined.

Taking out a common factor $G_{l'l}$, which is determined by the total
decay rate, the coupling constants $g^n_{\epsilon\omega}$
are normalized to \cite{FGJ:86}
\bel{eq:normalization}
1 = \sum_{n,\epsilon,\omega}\, |g^n_{\epsilon\omega}/N^n|^2 \, ,
\ee
where 
$N^n %\equiv \mbox{\rm max}(|g^n_{\epsilon\omega }|) 
=2$, 1,
$1/\protect\sqrt{3} $ for $n =$ S, V, T.
In the SM, $g^V_{LL}  = 1$  and all other
$g^n_{\epsilon\omega} = 0 $.

The couplings $g^n_{\epsilon\omega}$ can be investigated through the
measurement of the final charged--lepton distribution 
%(with known initial and/or final polarizations) 
and with the inverse decay
$\nu_{l'} l\to l' \nu_l$. 
For $\mu$ decay, where precise measurements of the polarizations of
both $\mu$ and $e$ have been performed, 
there exist \cite{PDG:96}
stringent upper bounds on the couplings involving right--handed helicities.
These limits show nicely 
that the bulk of the $\mu$--decay transition amplitude is indeed of
the predicted V$-$A type:
$|g^V_{LL}|  > 0.96$ (90\% CL).
Improved measurements of the $\mu$ decay parameters will
be performed at PSI and TRIUMPH \cite{guill}.

%%%%%%%%%%%  FIGURES Mu and Tau Couplings %%%%%%%%%%%%%
\begin{figure}[tbh] %p]
\vfill
\centerline{
\begin{minipage}[t]{.47\linewidth}  %\centering
%\centerline{\framebox(50,20){FIGURE}}
\centerline{\epsfxsize =5.64cm \epsfbox{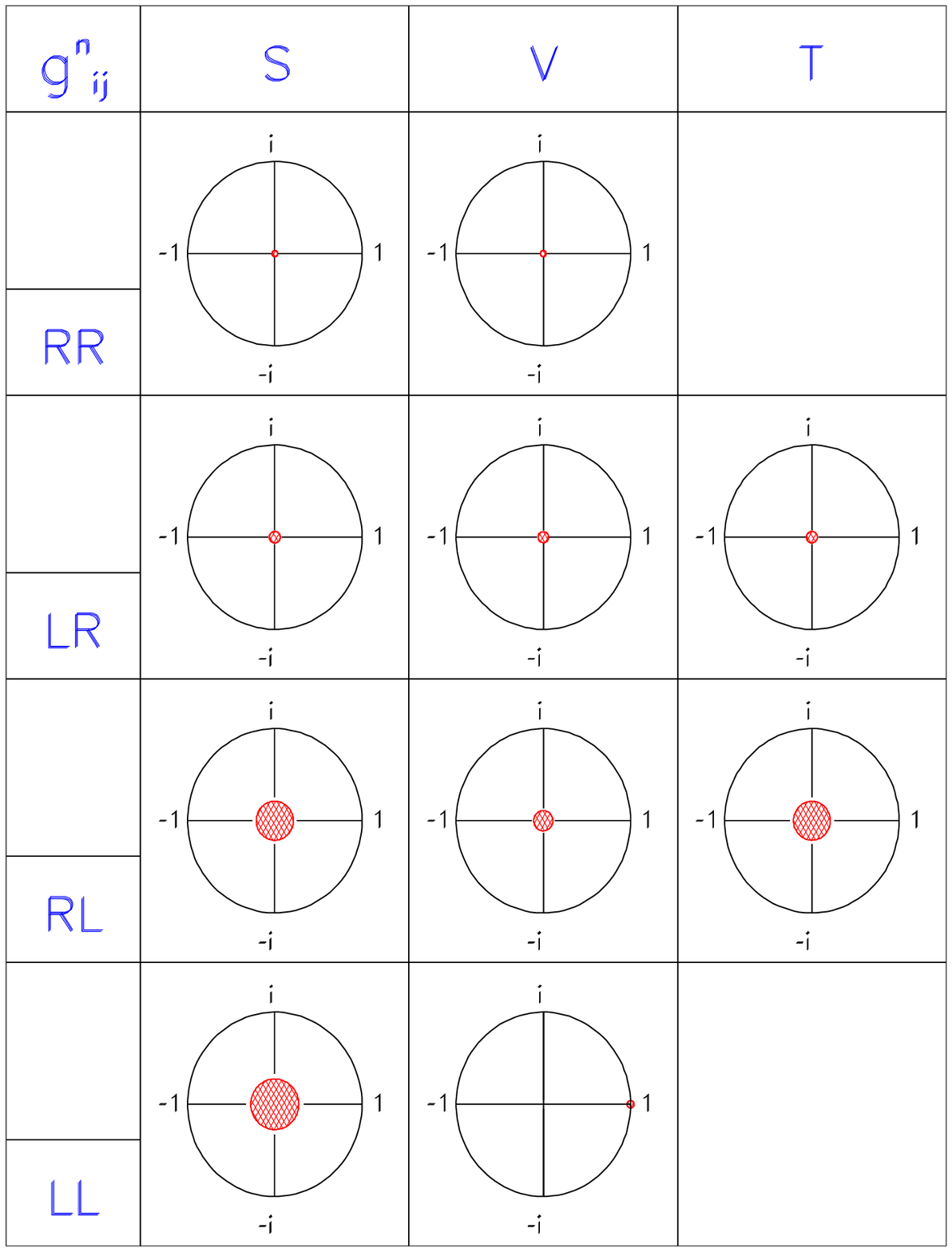}}
%\vspace{-1.5cm}
\caption{90\% CL experimental limits  \protect\cite{PDG:96}
for the normalized $\mu$--decay couplings
$g'^n_{\epsilon\omega }\equiv g^n_{\epsilon\omega }/ N^n$.
(Taken from Ref.~\protect\citen{LR:95}). \hfill\hfil }
\label{fig:mu_couplings}
\end{minipage}
\hspace{0.72cm}
\begin{minipage}[t]{.47\linewidth}  %\centering
%\centerline{\framebox(50,20){FIGURE}}
\centerline{\epsfxsize =5.64cm \epsfbox{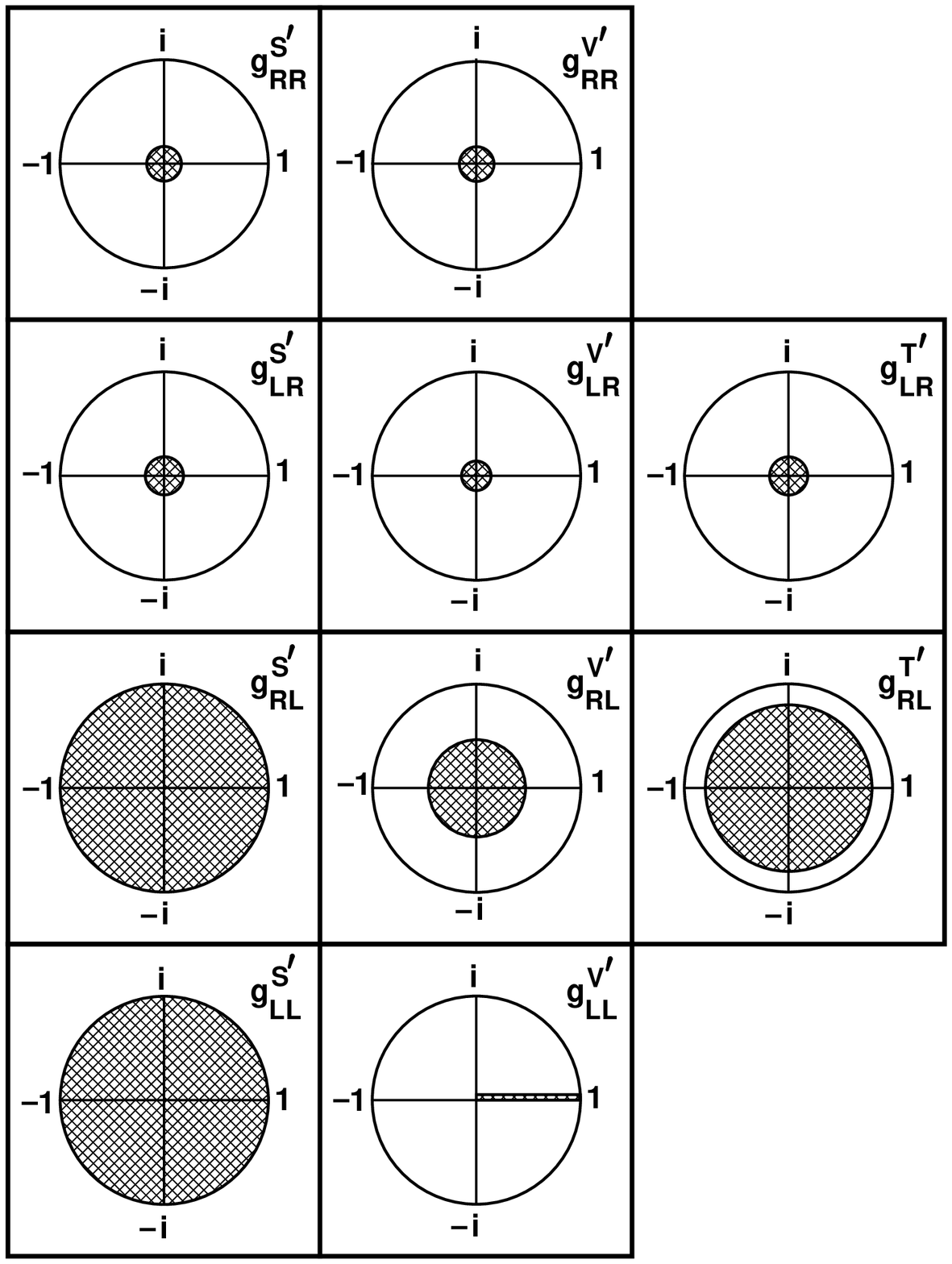}}
%\vspace{-1.5cm}
\caption{90\% CL experimental limits \protect\cite{cleo:97a}
for the normalized $\tau$--decay couplings
$g'^n_{\epsilon\omega }\equiv g^n_{\epsilon\omega }/ N^n$,
assuming $e/\mu$ universality. \hfill\hfil }
\label{fig:tau_couplings}
\end{minipage}
}
\vfill
\end{figure}
%%%%%%%%%%%%%%%%%%%%% END FIGURES %%%%%%%%%%%%%%%%%%%%%%%%%%

The $\tau$--decay experiments are starting to provide
useful information on the decay structure.
Figure \ref{fig:tau_couplings} shows the most recent limits
obtained by CLEO \cite{cleo:97a}.
The measurement of the $\tau$ polarization allows to bound those couplings
involving an initial right--handed lepton; however, information on the
final charged--lepton polarization is still lacking. Moreover,
the measurement of the inverse decay
$\nu_\tau l\to\tau\nu_l$, needed to separate the $g^S_{LL}$ and
$g^V_{LL}$ couplings, looks far out of reach.

\section{Neutral--Current Couplings}
\label{sec:nc}

In the SM, all fermions with equal electric charge have identical
couplings to the $Z$ boson:
\bel{eq:L_nc}
\cL_{\mbox{\rms NC}}^Z \, = \, { g \over 2 \cos{\theta_W}} \,
     Z_\mu \,\sum_l \bar f \gamma^\mu (v_f - a_f \gamma_5) f \, ,
\ee
where 
\be
v_f = T_3^f (1-4|Q_f|\sin^2{\theta_W})\ ,
\qquad\qquad
a_f=T_3^f\, .
\ee
These neutral current couplings have been precisely
tested at LEP and SLC \cite{lepewwg:97}.

\subsection{Tree--Level Phenomenology}
\label{subsec:tree-level}

The gauge sector of the SM is fully described in terms of only
four parameters:
$g$, $g'$, 
and the two constants characterizing the scalar potential.
We can trade these
parameters by $\alpha$, $\theta_W$, $M_W$ and $M_H$.
Alternatively, one can choose as free parameters
$\alpha$, $M_Z$, $G_F$ and $M_H$; this has the advantage of using the 3 
most precise experimental determinations to fix the 
interaction.

Taking as inputs Eqs.~\eqn{eq:alpha}, \eqn{eq:gf} and \cite{lepewwg:97}
\bel{eq:SM_inputs}
M_Z  =  (91.1867\pm 0.0020)\,\mbox{\rm GeV} \, , 
\ee
the relations
\bel{eq:A_def}
M_W^2 s_W^2  =  {\pi\alpha\over\sqrt{2} G_F}\equiv
\Omega =  [(37.2802\pm 0.0003)\,\mbox{\rm GeV}]^2 ,
\ee
\bel{eq:sW_def}
s_W^2  =  1 - {M_W^2\over M_Z^2}\, ,
\ee
determine $s_W \equiv \sin{\theta_W}$ and $M_W$:
\bel{eq:M_W_det}
M_W \, =\,  {M_Z\over\sqrt{2}}\,
\left\{1 + \sqrt{1-{4\Omega\over M_Z^2}}\right\}^{1/2}
 = \, 80.94\, \mbox{\rm GeV} , 
\ee 
\bel{eq:s_W_det} 
s_W^2  \, =\,  {1\over 2}\, \left\{1 -\sqrt{1-{4\Omega\over M_Z^2}}\right\}
\, =\,  0.2122 \, .
\ee
The predicted $W$ mass is in 
reasonable agreement with the measured value \cite{lepewwg:97}, 
$M_W = 80.43\pm 0.08$ GeV.

At tree level, the partial decay widths of the $Z$ boson can be
easily computed:
\bel{eq:Z_width}
\Gamma\left[ Z\to \bar f f\right]  =  
{G_F M_Z^3\over 6\pi\sqrt{2}} \, \left(|v_f|^2 + |a_f|^2\right)\, 
N_f  
% =  0.3318\, \left(|v_f|^2 + |a_f|^2\right)\, N_f\, \,\mbox{\rm GeV} 
\, ,
\ee
where $N_l=1$ and $N_q=N_C$.
Summing over all possible final fermion pairs, one predicts
the total width
$\Gamma_Z=2.474$ GeV, to be compared
with the experimental value \cite{lepewwg:97}
$\Gamma_Z=(2.4948\pm 0.0025)$ GeV.
The leptonic decay widths of the $Z$ are predicted to be 
$\Gamma_l\equiv\Gamma(Z\to l^+l^-) = 
84.84$ MeV,
in agreement with the measured value
$\Gamma_l = (83.91\pm 0.10)$ MeV.

Other interesting quantities are the $Z$ decay width into invisible modes,
\bel{eq:Z_inv}
 {\Gamma_{\mbox{\rms inv}}\over\Gamma_l}\,\equiv\,
{N_\nu\,\Gamma(Z\to\bar\nu\nu)\over\Gamma_l}\, = \,
{N_\nu\over 2\, (|v_l|^2 + |a_l|^2)} \, = \, 5.866 \, ,
\ee
which is usually normalized to the (charged) leptonic width,
and the ratios
\beqn\label{eq:Z_had}
R_l &\!\!\!\equiv &\!\!\! {\Gamma(Z\to\mbox{\rm hadrons})\over\Gamma_l}
\, = \, {\sum_q N_q \, (|v_q|^2 + |a_q|^2) \over |v_l|^2 + |a_l|^2}
\, = \,
20.29 \, ,
\no\\ \label{eq:R_b_def}
R_b &\!\!\!\equiv &\!\!\! 
{\Gamma(Z\to\bar b b)\over \Gamma(Z\to\mbox{\rm hadrons})}
\, = \, {|v_b|^2 + |a_b|^2 \over \sum_q (|v_q|^2 + |a_q|^2)}
\, = \, 0.219 \, ,
\\ \label{eq:R_c_def}
R_c &\!\!\!\equiv &\!\!\! 
{\Gamma(Z\to\bar c c)\over \Gamma(Z\to\mbox{\rm hadrons})}
\, = \, {|v_c|^2 + |a_c|^2 \over \sum_q (|v_q|^2 + |a_q|^2)}
\, = \, 0.172 \, . \no
\eeqn
The comparison with the experimental values, shown  in
Table~\ref{tab:results}, is quite good.

%%%%%%%%%%%%%%%%%%%%%%%%%

Additional information can be obtained from the study of the
fermion--pair production process
\bel{eq:production}
e^+e^-\to\gamma,Z\to\bar f f \, .
\ee
For unpolarized $e^+$ and $e^-$ beams, the differential 
$e^+e^-\to \gamma,Z\to \bar f f$
cross-section can be written, at lowest order, as
\bel{eq:dif_cross}
{d\sigma\over d\Omega} =  {\alpha^2\over 8 s} \, N_f \,
         \left\{ A \, (1 + \cos^2{\theta})  + B\,  \cos{\theta}
     -  h_f \left[ C \, (1 + \cos^2{\theta})  + D \cos{\theta}
         \right] \right\} \! ,
\ee
where $h_f$ ($=\pm1$) is the helicity of the produced fermion $f$
and $\theta$ is 
the scattering angle between $e^-$ and $f$.
Here,
\be\label{eq:ABCD}\begin{array}{ccl}
A & = & 1 + 2 v_e v_f \,\mbox{\rm Re}(\chi)
 + \left(v_e^2 + a_e^2\right) \left(v_f^2 + a_f^2\right) |\chi|^2, 
\\
B & = & 4 a_e a_f \,\mbox{\rm Re}(\chi) + 8 v_e a_e v_f a_f  |\chi|^2  , 
\\
C & = & 2 v_e a_f \,\mbox{\rm Re}(\chi) + 2 \left(v_e^2 + a_e^2\right) 
  v_f a_f |\chi|^2 ,
\\
D & = & 4 a_e v_f \,\mbox{\rm Re}(\chi) + 4 v_e a_e \left(v_f^2 +
      a_f^2\right) |\chi|^2  , 
\\ \ea
\ee
and  $\chi$  contains the $Z$  propagator
\bel{eq:Z_propagator}
\chi \, = \, {G_F M_Z^2 \over 2 \sqrt{2} \pi \alpha }
     \,\, {s \over s - M_Z^2 + i s \Gamma_Z  / M_Z } \, . 
\ee

The coefficients $A$, $B$, $C$ and $D$ can be experimentally determined,
by measuring the total cross-section, the forward--backward asymmetry,
the polarization asymmetry and the forward--backward polarization
asymmetry, respectively:
$$
\sigma(s)  =  {4 \pi \alpha^2 \over 3 s } \, N_f \, A \, , 
\qquad\qquad
\cA_{\mbox{\rms FB}}(s) \equiv   {N_F - N_B \over N_F + N_B}
     =  {3 \over 8} {B \over A}\,  ,
$$
\be\label{eq:A_pol}
\cA_{\mbox{\rms Pol}}(s)  \equiv
{\sigma^{(h_f =+1)}
- \sigma^{(h_f =-1)} \over \sigma^{(h_f =+1)} + \sigma^{(h_f = -1)}}
\, = \,  - {C \over A} \, ,
\ee
$$
\cA_{\mbox{\rms FB,Pol}}(s)  \equiv  
{N_F^{(h_f =+1)} - 
N_F^{(h_f = -1)} - N_B^{(h_f =+1)} + N_B^{(h_f = -1)} \over
N_F^{(h_f =+1)} + N_F^{(h_f = -1)} + N_B^{(h_f =+1)} + N_B^{(h_f = -1)}}
\, = \, -{3 \over 8} {D \over A}\, . 
$$
$N_F$ and $N_B$ denote the number of $f$'s
emerging in the forward and backward hemispheres,
respectively, with respect to the electron direction.

For $s = M_Z^2$,
the real part of the $Z$ propagator vanishes
and the photon exchange terms can be neglected
in comparison with the $Z$--exchange contributions
($\Gamma_Z^2 / M_Z^2 \ll 1$). Eqs.~\eqn{eq:A_pol}
become then,
\beqn
\sigma^{0,f}  \equiv  \sigma(M_Z^2)  = 
 {12 \pi  \over M_Z^2 } \, {\Gamma_e \Gamma_f\over\Gamma_Z^2}\, ,
&& \qquad\;
\cA_{\mbox{\rms FB}}^{0,f}\equiv\cA_{FB}(M_Z^2) = {3 \over 4}
\cP_e \cP_f \, ,
\no\\ \label{eq:A_pol_Z}
\cA_{\mbox{\rms Pol}}^{0,f} \equiv
  \cA_{\mbox{\rms Pol}}(M_Z^2)  = \cP_f \, ,\quad
&&  \qquad
\cA_{\mbox{\rms FB,Pol}}^{0,f} \equiv 
\cA_{\mbox{\rms FB,Pol}}(M_Z^2)  =  {3 \over 4} \cP_e  \, ,\quad\quad
\eeqn
where 
$\Gamma_f$ is the $Z$ partial decay width to the $\bar f f$ final state, and
\bel{eq:P_f}
\cP_f \, \equiv \, { - 2 v_f a_f \over v_f^2 + a_f^2} 
\ee
is the average longitudinal polarization of the fermion $f$,
which only depends on the ratio of the vector and axial--vector couplings.
$\cP_f$ is a sensitive function of $\sin^2{\theta_W}$.

The measurement of the final polarization asymmetries can (only) be done for 
$f=\tau$, because the spin polarization of the $\tau$'s
is reflected in the distorted distribution of their decay products.
Therefore, $\cP_\tau$ and $\cP_e$ can be determined from a
measurement of the spectrum of the final charged particles in the
decay of one $\tau$, or by studying the correlated distributions
between the final products of both $\tau's$ \cite{ABGPR:92}.

With polarized $e^+e^-$ beams, one can also study the left--right
asymmetry between the cross-sections for initial left-- and right--handed
electrons.
At the $Z$ peak, this asymmetry directly measures 
the average initial lepton polarization, $\cP_e$,
without any need for final particle identification:
\bel{eq:A_LR}
\cA_{\mbox{\rms LR}}^0\,\equiv\, \cA_{\mbox{\rms LR}}(M_Z^2)
  \, = \, {\sigma_L(M_Z^2)
- \sigma_R(M_Z^2) \over \sigma_L(M_Z^2) + \sigma_R(M_Z^2)}
\, = \,  - \cP_e \,  .
\ee
SLD has also measured the left--right forward--backward asymmetry
for $b$ and $c$ quarks, which are only sensitive to the
final state couplings:
\bel{eq:A_FB_LR}
\cA_{\rms FB,LR}^{0,f}\,\equiv\,\cA_{\rms FB,LR}^{f}(M_Z^2) 
\, = \, -{3\over 4} \cP_f \, .
\ee
%

%%%%%%%%%%%%%%%%% Table %%%%%%%%%%%%%%%%%%%%%%%%%%%%%%
\begin{table}[htb]
\begin{center}
\caption{Comparison between tree--level SM predictions and
experimental \protect\cite{lepewwg:97} measurements. 
The third column includes
the main QED and QCD corrections.
The experimental value for $s_W^2$ refers to the
effective electroweak mixing angle in the charged--lepton sector,
%defined in Eq.~\protect\eqn{eq:bar_s_W_l}.
\protect{
$\sin^2{\theta^{\mbox{\rms lept}}_{\mbox{\rms eff}}}\equiv (1-v_l/a_l)/4$}.
\hfill 
\label{tab:results}}
\vspace{0.2cm}
\begin{tabular}{|c|c|c|c|}
\hline
Parameter & \multicolumn{2}{c|}{Tree--level prediction} & 
Experimental  
\\ \cline{2-3} & Naive & Improved & value \\ \hline  %\hline
$M_W$ \, (GeV) & 80.94 & 79.96 & $80.43\pm 0.08$
\\
$s_W^2$ & 0.2122 & 0.2311 & $0.23152\pm 0.00023$
\\
%$\Gamma_W$ \,\, (GeV) & 2.09 & 2.06 & $2.08\pm 0.07$\\
$\Gamma_Z$ \, (GeV) & 2.474 & 2.490 & $2.4948\pm 0.0025$
\\
%Br($W^-\to\bar\nu_l l^-$) \,\, (\%) & 11.1 & 10.8 & $10.76\pm 0.33$\\
$\Gamma_l$ \, (MeV) & 84.84 & 83.41 & $83.91\pm 0.10$
\\
$\Gamma_{\mbox{\rms inv}}/\Gamma_l$ & 5.866 & 5.966 & $5.960\pm 0.022$
\\
$R_l$ & 20.29 & 20.88 & $20.775\pm 0.027$
\\
$\sigma^0_{\mbox{\rms had}}$ \,\, (nb) & 42.13 & 41.38 & $41.486\pm 0.053$
\\
$\cA_{\mbox{\rms FB}}^{0,l}$ & 0.0657 & 0.0169 & $0.0171\pm 0.0010$
\\
$\cP_l$ & $-0.296$ & $-0.150$ & $-0.1505\pm 0.0023$
\\
%$\cA_{\mbox{\rms Pol}}^{0,\tau}$ & $-0.296$ & $-0.148$ & $-0.143\pm 0.010$\\
%${4\over 3}\cA^{0,\tau}_{\mbox{\rms FB,Pol}}$ & $-0.296$ & $-0.148$ & 
%$-0.135\pm 0.011$\\
%$-\cA_{\mbox{\rms LR}}^0$ & $-0.296$ & $-0.148$ & $-0.1637\pm 0.0075$\\
$\cA_{\mbox{\rms FB}}^{0,b}$ & 0.210 & 0.105 & $0.0984\pm 0.0024$
\\
$\cA_{\mbox{\rms FB}}^{0,c}$ & 0.162 & 0.075 & $0.0741\pm 0.0048$
\\
$\cP_b$ & $-0.947$ & $-0.936$ & $-0.900\pm 0.050$
\\
$\cP_c$ & $-0.731$ & $-0.669$ & $-0.650\pm 0.058$
\\
$R_b$ & 0.219 & 0.220 & $0.2170\pm 0.0009$
\\
$R_c$ & 0.172 & 0.170 & $0.1734\pm 0.0048$
\\ \hline
\end{tabular}
\end{center}
\end{table}
%%%%%%%%%%%%%%%% End Table %%%%%%%%%%%%%%%%%%%%%%%%%%%%

Using the value of the weak mixing angle determined in Eq.~\eqn{eq:s_W_det},
one gets the predictions shown in the second column of 
Table~\ref{tab:results}.
The comparison with the experimental measurements
looks reasonable for the total hadronic cross-section 
$\sigma^0_{\mbox{\rms had}} \equiv \sum_q \, \sigma^{0,q}$;
however, all leptonic asymmetries disagree with
the measured values by several standard deviations.
As shown in the table, the same happens with the 
heavy--flavour forward--backward asymmetries
$\cA_{\mbox{\rms FB}}^{0,b/c}$,
which compare very badly with the experimental measurements;
the agreement is however better for $\cP_{b/c}$.

Clearly, the problem with the asymmetries is their high sensitivity
to the input value of $\sin^2{\theta_W}$;
specially the ones involving the leptonic vector coupling
$v_l = (1 - 4 \sin^2{\theta_W})/2$. Therefore, they are an
extremely good window into higher--order electroweak corrections.

\subsection{Important QED and QCD Corrections}
\label{subsec:QED_QCD_corr}

Before trying to analyze the relevance of higher--order electroweak 
contributions, it is instructive to consider the 
well--known QED and QCD corrections.

The photon propagator gets vacuum polarization corrections, induced by
virtual fermion--antifermion pairs. Their effect can be
taken into account through a redefinition of the QED coupling,
which depends on the energy scale of the process;
the resulting effective coupling $\alpha(s)$
is called the QED {\it running coupling}.
The fine structure constant in Eq.~\eqn{eq:alpha} is measured
at very low energies; it corresponds to $\alpha(m_e^2)$.
However, at the $Z$ peak, we should rather use $\alpha(M_Z^2)$.
The long running from $m_e$ to $M_Z$ gives rise to a sizeable
correction \cite{ADH:97,EJ:95}:
\bel{eq:QED_corr}
\alpha^{-1}\equiv\alpha(m_e^2)^{-1}\,\Longrightarrow\,
\alpha(M_Z^2)^{-1}\equiv \alpha^{-1} \, (1-\Delta\alpha) = 
128.896\pm  0.090\, .
\ee
The quoted uncertainty arises from the light--quark contribution,
which is estimated from $\sigma(e^+e^-\to\mbox{\rm hadrons})$ and 
$\tau$--decay data. 

This running effect generates an important change in Eq.~\eqn{eq:A_def}.
Since $G_F$ is measured at low energies, while $M_W$ is a
high--energy parameter, the relation between both quantities is clearly
modified by vacuum--polarization contributions:
\bel{eq:barA_def}
M_W^2 s_W^2 \, = \, {\pi\alpha(M_Z^2)\over\sqrt{2} G_F}
\, = \, {\Omega\over 1 -\Delta\alpha}
\,\equiv\,
\overline\Omega\, = \, [38.439\,\mbox{\rm GeV}]^2 ,
\ee
Changing $\Omega$ by $\overline\Omega$ in Eqs.~\eqn{eq:M_W_det} 
and \eqn{eq:s_W_det},
one gets the corrected predictions:
\bel{eq:new_pred}
M_W\, = \, 79.96\,\mbox{\rm GeV} , \qquad\qquad\qquad
s^2_W \, = \, 0.2311 \, .
\ee

So far, we have treated quarks and leptons on an equal footing.
However, quarks are strong--interacting particles. 
The gluonic corrections to the $Z\to\bar q q$ decays 
%and $W^-\to\bar u_i d_j$ 
can be directly incorporated
into the formulae given before, by taking an effective number of
colours:
\bel{N_C_eff}
N_q \, = \, 
N_C\,\left\{ 1 + {\alpha_s\over\pi} + \ldots\right\}\,
\approx\, 3.12 \, ,
\ee
where we have used $\alpha_s(M_Z^2)\approx 0.12\, $.
Note that the strong coupling also {\it runs}; one should then use
the value of $\alpha_s$ at $s=M_Z^2$.

The third column in Table~\ref{tab:results} shows the numerical impact
of these  QED and QCD corrections. In all cases, the comparison with
the data gets improved. However, it is in the asymmetries where the
effect gets more spectacular. Owing to the high sensitivity to $s^2_W$,
the small change in the value of the weak mixing angle generates
a huge difference of about a factor of 2 in the predicted
asymmetries.
The agreement with the experimental values is now very good.

\subsection{Higher--Order Electroweak Corrections}
\label{subsec:nc-loop}

%%%%%%%%%%%%%%%%   FIGURE LEP %%%%%%%%%%%%%%%%
\begin{figure}[tbh]
%\centerline{\framebox(50,20){FIGURE}}
\centerline{\epsfxsize =10cm \epsfbox{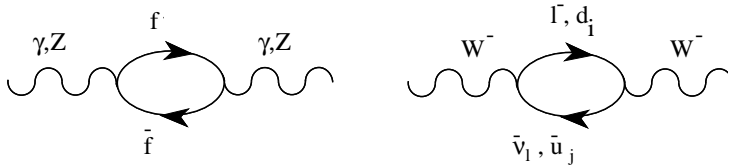}}
\vspace{-0.2cm}
\caption{Gauge--boson self-energies (vacuum polarization corrections).
\hfill\hfil }
\label{fig:oblique}
\end{figure}
%%%%%%%%%%%%%%%%%%%%% END FIGURE %%%%%%%%%%%%%%%%%%%%%%%%%%

 Initial-- and final--state photon radiation is by far the
most important numerical correction. One has in addition the contributions
coming from photon exchange between the fermionic lines. 
All these QED corrections are to a large extent dependent on the detector and
the experimental cuts, because of the infra-red problems associated with
massless photons (one needs to define, for instance, the minimun photon energy
which can be detected). Therefore, these effects are usually estimated with
Monte Carlo programs and subtracted from the data.
Notice that in the decay $\mu^-\to e^- \bar\nu_e \nu_\mu$,
the QED corrections are already partly included in the definition of $G_F$
thus, one should take care of subtracting those
corrections already incorporated in Eq.~\eqn{eq:qed_corr}.

More interesting are the so--called {\it oblique} corrections,
gauge--boson self-energies induced by vacuum polarization diagrams,
which are {\it universal} (process independent). 
We have already seen the important role of the photon self-energy.
In the case of the $W^\pm$ and the $Z$, these corrections are sensitive
to heavy particles (such as the top) running along the loop \cite{VE:77}.

In QED, the 
vacuum polarization contribution of a heavy fermion pair
is suppressed by inverse powers of the fermion mass.
At low energies ($s<<m_f^2$), 
the information on the heavy fermions is then lost.
This {\it decoupling} of the heavy fields happens in theories
like QED and QCD,
with only vector couplings and an exact gauge symmetry \cite{AC:75}.
The SM involves, however, a broken chiral gauge symmetry. This
has the very interesting implication of avoiding the decoupling
theorem \cite{AC:75}, offering the possibility to be sensitive to 
heavy particles which cannot be kinematically accessed.

The $W^\pm$ and $Z$ self-energies induced by a heavy top,
i.e. $W^-\to \bar t b \to W^-$ and $Z\to\bar t t \to Z$,
generate contributions 
which increase quadratically with the top mass \cite{VE:77}.
The leading $m_t^2$ contribution to the $W^\pm$ propagator 
amounts to a $-3\% $  correction to the relation \eqn{eq:A_def}
between $G_F$ and $M_W$.

Owing to an accidental $SU(2)_C$ symmetry of the scalar sector
(the so--called custodial symmetry), the
virtual production of Higgs particles does not generate any
$m_H^2$ dependence at one loop (Veltman screening \cite{VE:77}).
The dependence on the Higgs mass is only logarithmic.
The numerical size of the correction induced on \eqn{eq:A_def}
is $-0.3\% $ ($+1\% $) for $m_H=60$ (1000) GeV.

%%%%%%%%%%%%%%%%   FIGURE LEP %%%%%%%%%%%%%%%%
\begin{figure}[tbh]
%\centerline{\framebox(50,20){FIGURE}}
\centerline{\epsfxsize =9cm \epsfbox{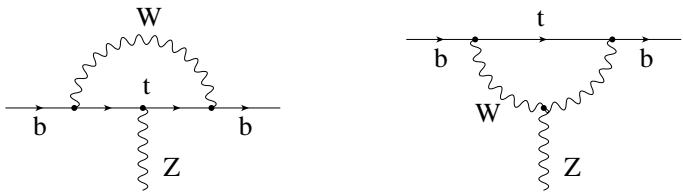}}
\vspace{-0.2cm}
\caption{$m_t$--dependent corrections to the $Z\bar b b$ vertex.
\hfill\hfil }
\label{fig:zbb}
\end{figure}
%%%%%%%%%%%%%%%%%%%%% END FIGURE %%%%%%%%%%%%%%%%%%%%%%%%%%

The vertex corrections to the different couplings
are {\it non-universal} and usually smaller than the oblique contributions.
There is one interesting exception, the $Z \bar bb$ vertex, which is sensitive
to the top quark mass \cite{BPS:88}.
The $Z\bar f f$ vertex gets 1--loop corrections where a virtual
$W^\pm$ is exchanged between the two fermionic legs. Since, the $W^\pm$
coupling changes the fermion flavour, the decays
$Z\to \bar d d, \bar s s, \bar b b\ $ get contributions with a top quark
in the internal fermionic lines.
These amplitudes are suppressed by a small quark--mixing factor 
$|V_{td_i}|^2$,
except for the $Z\to\bar b b$ vertex because $|V_{tb}|\approx 1$.

The explicit calculation \cite{BPS:88,ABR:86,BH:88,LS:90}
shows the presence of hard $m_t^2$ corrections to
the $Z\to\bar b b$ vertex. This effect can be easily understood
\cite{BPS:88}
in non-unitary gauges where the unphysical charged scalar $\phi^{(\pm)}$
is present. The Yukawa couplings of the charged scalar to fermions
are proportional to the fermion masses; therefore, the exchange
of a virtual $\phi^{(\pm)}$ gives rise to a $m_t^2$ factor.
In the unitary gauge, the charged scalar has been {\it eaten} by the
$W^\pm$ field; thus, the effect comes now from the exchange of a
longitudinal $W^\pm$,
with terms proportional to $q^\mu q^\nu$ in the propagator that
generate fermion masses.

Since the $W^\pm$ couples only to left--handed fermions, the induced
effect is the same on the vector
and axial--vector $Z\bar b b$ couplings.
It amounts \cite{BPS:88} to a $-1.5\% $ correction of $\Gamma(Z\to\bar b b)$.

The {\it non-decoupling} present in the
$Z\bar b b$ vertex is quite different from the one happening in
the boson self-energies. 
The vertex correction does not have any dependence with the
Higgs mass. Moreover,
while any kind of new heavy particle,
coupling to the gauge bosons, would contribute to the $W^\pm$ and $Z$
self-energies, possible new--physics contributions to the
$Z\bar b b$ vertex are much more restricted and, in any case,
different.
Therefore, an independent experimental test of the two effects
is very valuable in order to disentangle possible
new--physics contributions from the SM corrections.

The remaining quantum corrections are rather small.
Box diagrams with two gauge--boson exchanges
give a very small contribution at the $Z$ peak,
because they are non resonant (they do not have an on-shell $Z$ propagator). 
However, the box correction to the
decay $\mu^-\to e^- \bar\nu_e \nu_\mu$ is not negligible.
The exchange of a Higgs particle between two fermionic lines
is irrelevant, because the amplitude is suppressed
by the product of the two fermionic masses.

\subsection{Lepton Universality}

%%%%%%%%%%%%%  Table %%%%%%%%%%%%%%%%%%%%%%%%
\begin{table}[htb]
\centering
\caption{Measured values \protect\cite{lepewwg:97}
of $\Gamma_l\equiv\Gamma(Z\to l^+l^-)$
and the leptonic forward--backward asymmetries.
The last column shows the combined result 
(for a massless lepton) assuming lepton universality. \hfill
\label{tab:LEP_asym}}
\vspace{0.2cm}
\begin{tabular}{|c|ccc|c|}
\hline
& $e$ & $\mu$ & $\tau$ & $l$ 
\\ \hline
$\Gamma_l$ \, (MeV) & $83.94\pm 0.14$
& $83.84\pm 0.20$ & $83.68\pm 0.24$ & $83.91\pm 0.10$
\\
$\cA_{\mbox{\rms FB}}^{0,l}$ \, (\%) & $1.60\pm 0.24$
& $1.63\pm 0.14$ & $1.92\pm 0.18$ & $1.71\pm 0.10$
\\ \hline
\end{tabular}
%\end{table}
%%%%%%%%%%%%%  End Table %%%%%%%%%%%%%%%%%%%%%%%%
%%%%%%%%%%%%%  Table %%%%%%%%%%%%%%%%%%%%%%%%
%\begin{table}
\centering
\caption{Measured values \protect\cite{lepewwg:97}
of the leptonic polarization asymmetries.}
\label{tab:pol_asym}
\vspace{0.2cm}
\begin{tabular}{|c|c|c|c|}
\hline
$-\cA_{\mbox{\rms Pol}}^{0,\tau} = -\cP_\tau$ &
$-{4\over 3}\cA^{0,\tau}_{\mbox{\rms FB,Pol}} = -\cP_e$ &
$\cA_{\mbox{\rms LR}}^0 = -\cP_e$
& $\!\{{4\over 3}\cA_{\mbox{\rms FB}}^{0,l}\}^{1/2} = -P_l\! $
\\ \hline
$0.1410\pm 0.0064$ & $0.1399\pm 0.0073$ & $0.1547\pm 0.0032$
& $0.1510\pm 0.0044$
\\ \hline
\end{tabular}
\end{table}
%%%%%%%%%%%%% End Tables %%%%%%%%%%%%%%%%%%%%%

%%%%%%%%%%%%%%% Table %%%%%%%%%%%%%%%%%
\begin{table}[bth]
\centering
\caption{
Effective vector and axial--vector lepton couplings
derived from LEP and SLD  \protect\cite{lepewwg:97}.
\label{tab:nc_measured}}
\vspace{0.2cm}
\begin{tabular}{|c|c|c|}     
\hline
& \multicolumn{2}{c|}{Without Lepton Universality}\\ \cline{2-3}
& LEP & LEP + SLD\\ \hline
$v_e$ & $-0.0367\pm 0.0015$ & 
        $-0.03844 \pm 0.00071$
\\
$v_\mu$ & $-0.0374\pm 0.0036$ & 
          $-0.0358 \pm 0.0032$
\\
$v_\tau$ & $-0.0367\pm 0.0015$ & 
           $-0.0365 \pm 0.0015$
\\
$a_e$ & $-0.50123\pm 0.00044$ & 
        $-0.50111 \pm 0.00043$
\\
$a_\mu$ & $-0.50087\pm 0.00066$ & 
          $-0.50098 \pm 0.00065$
\\
$a_\tau$ & $-0.50102\pm 0.00074$ & 
           $-0.50103 \pm 0.00074$
\\ \hline  
$v_\mu/v_e$ & $1.02\pm 0.12$ & 
              $\phantom{-}0.932\pm 0.087$
\\
$v_\tau/v_e$ & $0.998\pm 0.060$ & 
               $\phantom{-}0.949\pm 0.044$
\\
$a_\mu/a_e$ & $0.9993\pm 0.0017$ & 
              $\phantom{-}0.9997\pm 0.0016$
\\
$a_\tau/a_e$ & $0.9996\pm 0.0018$ & 
               $\phantom{-}0.9998\pm 0.0018$
\\ \hline
& \multicolumn{2}{c|}{With Lepton Universality}\\ \cline{2-3}
& LEP & LEP + SLD \\ \hline
$v_l$ & $-0.03681\pm 0.00085$ & 
        $-0.03793 \pm 0.00058$
\\
$a_l$ & $-0.50112\pm 0.00032$ & 
        $-0.50103 \pm 0.00031$
\\
$a_\nu=v_\nu$ & $+ 0.50125\pm 0.00092$ & 
                $+0.50125\pm 0.00092$ 
\\ \hline
\end{tabular}
\end{table}
%%%%%%%%%%%% End Table %%%%%%%%%%%%%%%%

Tables~\ref{tab:LEP_asym} and \ref{tab:pol_asym}
show the present experimental results
for the leptonic $Z$ decay widths and asymmetries.
The data are in excellent agreement with the SM predictions
and confirm the universality of the leptonic neutral couplings.\footnote{
%%%%%%%%%%
A small 0.2\% difference between $\Gamma_\tau$ and $\Gamma_{e,\mu}$
is generated by the $m_\tau$ corrections.}
%%%%%%%%%%%
There is however a small $1.9\sigma$ discrepancy between the
$\cP_e$ values obtained \cite{lepewwg:97} from 
$\cA^{0,\tau}_{\mbox{\rms FB,Pol}}$ 
and $\cA_{\mbox{\rms LR}}^0$.
The average of the two $\tau$ polarization measurements,
$\cA_{\mbox{\rms Pol}}^{0,\tau}$ and
${4\over 3}\cA^{0,\tau}_{\mbox{\rms FB,Pol}}$,
results in $\cP_l = -0.1406\pm 0.0048$ which disagrees
with the $\cA^0_{LR}$ measurement at the $2.4\sigma$ level.
Assuming lepton universality, 
the combined result from all leptonic asymmetries gives
\bel{eq:average_P_l}
\cP_l = - 0.1505\pm 0.0023 
\qquad (\chi^2/\mbox{\rm d.o.f.}  = 6.0/2) \ .
\ee

The measurement of $\cA_{\mbox{\rms Pol}}^{0,\tau}$ and
$\cA^{0,\tau}_{\mbox{\rms FB,Pol}}$ assumes that the $\tau$ decay
proceeds through the SM charged--current interaction.
A more general analysis should take into account the fact that the
$\tau$ decay width depends on the product $\xi\cP_\tau$ 
where $\xi$ ($=1$ in the SM)
is the corresponding Michel parameter in leptonic decays, or
the equivalent quantity $\xi_h$  ($=h_{\nu_\tau}$) in the semileptonic 
modes.
A separate measurement of $\xi$ and $\cP_\tau$ has been performed by
ALEPH \cite{ALEPH:94} ($\cP_\tau = -0.139\pm 0.040$)
and L3 \cite{L3:96} ($\cP_\tau = -0.154\pm 0.022$),
using the correlated distribution of the $\tau^+\tau^-$ decays.

%%%%%%%%%%%%%%%%   FIGURE LEP %%%%%%%%%%%%%%%%
\begin{figure}[bth] %p]
%\centerline{\framebox(50,20){FIGURE}}
\centerline{\epsfxsize =7cm \epsfbox{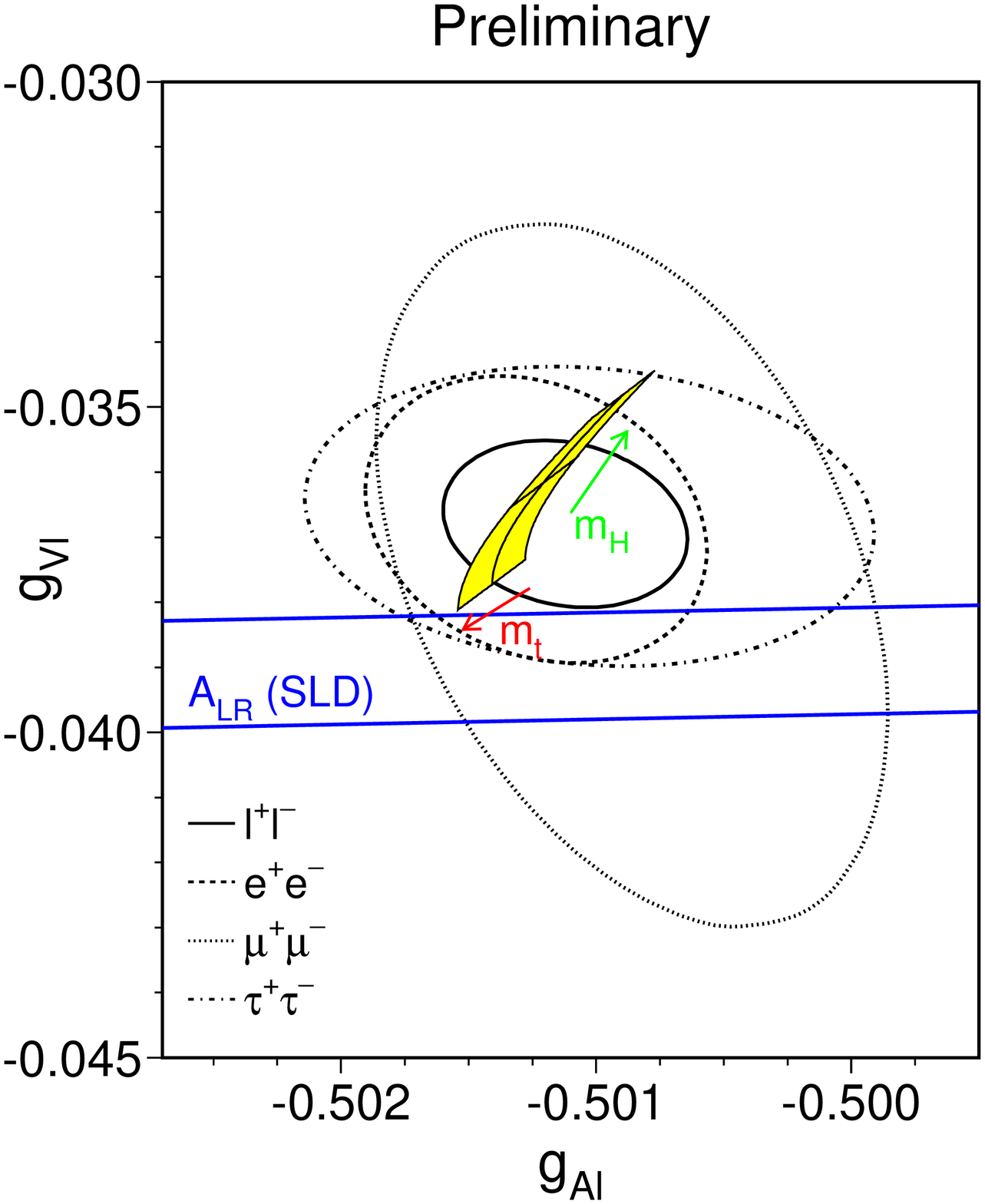}}
\vspace{-0.3cm}
\caption{68\% probability contours in the $a_l$-$v_l$ plane
from LEP measurements \protect\cite{lepewwg:97}. 
The solid contour assumes lepton universality. 
Also shown is the $1\sigma$ band resulting from the
$\protect\cA_{\mbox{\protect\rms LR}}^0$ measurement at SLD. 
The shaded region corresponds to the SM prediction
for \protect{$m_t= 175.6\pm 5.5$} GeV and
\protect{$m_H = 300^{+700}_{-240}$} GeV.
The arrows point in the direction of increasing
$m_t$ and $m_H$ values. \hfill\hfil }
\label{fig:gagv}
\end{figure}
%%%%%%%%%%%%%%%%%%%%% END FIGURE %%%%%%%%%%%%%%%%%%%%%%%%%%

The combined analysis \cite{lepewwg:97} of all leptonic observables 
from LEP and SLD ($\cA_{\mbox{\rms LR}}^0$) results
in the effective vector and axial--vector couplings given in
Table~\ref{tab:nc_measured}.\footnote{
%%%%%%%%%%% Footnote %%%%%%%%%%%%%%%%%
The asymmetries determine two possible solutions for
$|v_l/a_l|$. The ambiguity can be solved with lower--energy data
or through the measurement of the transverse spin--spin
correlation \protect\cite{BPR:91} of the two $\tau$'s in
$Z\to\tau^+\tau^-$, which requires \protect\cite{ALEPH:97b,DELPHI:97}
$|v_l/a_l|<<1$.
The signs of $v_l$ and $a_l$ are fixed by requiring $a_e<0$.}
%%%%%%%%%%%%%%%%%% End Footnote %%%%%%
The corresponding 68\% probability contours in the $a_l$--$v_l$ plane 
are shown in Figure~\ref{fig:gagv}.
The measured ratios of the $e$, $\mu$ and $\tau$ couplings
provide a test of charged--lepton universality in the neutral--current 
sector.

The neutrino couplings can be determined from the invisible 
$Z$--decay width, by assuming three identical neutrino generations
with left--handed couplings (i.e., $v_\nu=a_\nu$), 
and fixing the sign from neutrino scattering 
data \cite{CHARMII:94}.
The resulting experimental value \cite{lepewwg:97},
given in Table~\ref{tab:nc_measured},
is in perfect agreement with the SM.
Alternatively, one can use the SM prediction for 
$\Gamma_{\mbox{\rms inv}}/\Gamma_l$
to get a determination of the number of (light) neutrino flavours
\cite{lepewwg:97}:
\be
N_\nu = 2.993\pm 0.011 \, .
\ee
The universality of the neutrino couplings has been tested
with $\nu_\mu e$ scattering data, which fixes \cite{CHARMII:94b}
the $\nu_\mu$ coupling to the $Z$: \ 
$v_{\nu_\mu} =  a_{\nu_\mu} = 0.502\pm 0.017$.

%%%%%%%%%%%%%  Table %%%%%%%%%%%%%%%%%%%%%%%%
\begin{table}
\centering
\caption{Determinations \protect\cite{lepewwg:97}
of $\cP_b$ and $\cP_c$ from LEP data alone
(using $\cP_l = - 0.1461\pm 0.0033$),
%the LEP average for $\cP_l$), 
from SLD data alone,
and from LEP + SLD data (using $\cP_l = - 0.1505\pm 0.0023$)
%the LEP + SLD average for $\cP_l$)
assuming lepton universality. \hfill\hfil}
\label{tab:P_b_c}
\vspace{0.2cm}
\begin{tabular}{|c|c|c|c|}
\hline
& LEP & SLD & LEP + SLD
%\\ & ($\cP_l = - 0.1461\pm 0.0033$) && ($\cP_l = - 0.1505\pm 0.0023$)
\\ \hline  %\hline
$\cP_b$ & $-0.897\pm 0.030$ & $-0.900\pm 0.050$ & $-0.877\pm 0.023$ 
\\
$\cP_c$ & $-0.674\pm 0.046$ & $-0.650\pm 0.058$ & $-0.653\pm 0.037$ 
\\ \hline
\end{tabular}
\end{table}
%%%%%%%%%%%%% End Tables %%%%%%%%%%%%%%%%%%%%%

Using the measured value of $\cP_l$, one can extract $\cP_b$ and
$\cP_c$ from the forward--backward heavy--flavour asymmetries,
measured at LEP. The resulting values are shown in
Table~\ref{tab:P_b_c}, together with the direct SLD
determinations through $\cA^{0,b/c}_{\mbox{\rms FB,LR}}$,
and the combination of LEP and SLD measurements.
The LEP results are in excellent agreement with SLD,
and in reasonable agreement with the SM predictions
($\cP_b = -0.935$, $\cP_c = -0.668$).
However the combined LEP + SLD determination of $|\cP_b|$
is about $2.5 \sigma$ below the SM.
This discrepancy results from the sum of three different effects:
the LEP measurement of $\cA^{0,b}_{\mbox{\rms FB}}$ is low
($2.0\sigma$) compared with the SM; the SLD measurement of
$|\cP_b|$ is also slightly lower ($0.7\sigma$); 
and $\cA^0_{\mbox{\rms LR}}$ is high ($2.4 \sigma$) compared to the SM.

%%%%%%%%%%%%%  Table %%%%%%%%%%%%%%%%%%%%%%%%
\begin{table}[htb]
\centering
\caption{Comparison of several determinations \protect\cite{lepewwg:97} of
\protect{$\sin^2{\theta^{\mbox{\rms lept}}_{\mbox{\rms eff}}}$}.}
\label{tab:s2_W}
\vspace{0.2cm}
\begin{tabular}{|c|c|c|c|c|}
\hline
& $\sin^2{\theta^{\mbox{\rms lept}}_{\mbox{\rms eff}}}$ & Average & 
Cumul. Average & $\chi^2/\mbox{\rm dof}$
\\ \hline  %\hline
$\cA^{0,l}_{\mbox{\rms FB}}$ & $0.23102 \pm 0.00056$ &&&
\\
$\cP_\tau$ & $0.23228 \pm 0.00081$ &&&
\\
$\cP_e$ & $0.23243 \pm 0.00093$ & $0.23162 \, (41)$  % \pm 0.00041$
& $0.23162 \pm 0.00041$ & $2.6/2$
\\ \hline
$\cA^{0,b}_{\mbox{\rms FB}}$ & $0.23236 \pm 0.00043$ &&&
\\
$\cA^{0,c}_{\mbox{\rms FB}}$ & $0.23140 \pm 0.00111$ &
$0.23223 \, (40)$  % \pm 0.00040$ 
& $0.23194 \pm 0.00029$ & $4.3/4$
\\ \hline
$\langle Q_{\mbox{\rms FB}}\rangle$ & $0.2322 \pm 0.0010$ &
$0.2322 \, (10)$  %\pm 0.0010$ 
& $0.23196 \pm 0.00028$ & $4.4/5$
\\ \hline
$\cA^0_{\mbox{\rms LR}}$ & $0.23055 \pm 0.00041$ &
$0.23055 \, (41)$  %\pm 0.00041$ 
& $0.23152 \pm 0.00023$ & $12.5/6$
\\ \hline
\end{tabular}
\end{table}
%%%%%%%%%%%%% End Tables %%%%%%%%%%%%%%%%%%%%%

Assuming lepton universality,
the measured leptonic asymmetries can be used to obtain the
effective electroweak mixing angle in the charged--lepton sector,
defined as: 
\bel{eq:bar_s_W_l}
\sin^2{\theta^{\mbox{\rms lept}}_{\mbox{\rms eff}}} \equiv
{1\over 4}   \left( 1 - {v_l\over a_l}\right)  
\, .
\ee
One can also include the information provided by the hadronic
asymmetries, if the hadronic couplings are assumed to be given
by the SM; this is justified by the smaller sensitivity of
$\cP_{b/c}$ to higher--order corrections.
The different determinations \cite{lepewwg:97} 
of $\sin^2{\theta^{\mbox{\rms lept}}_{\mbox{\rms eff}}}$
and their combination
are shown in Table~\ref{tab:s2_W}.

\subsection{SM Electroweak Fit}

Including the full SM predictions at the 1--loop level,
the $Z$ measurements can be used to obtain information on
the SM parameters.
The high accuracy of the present data provides compelling evidence
for the pure weak quantum corrections, beyond the main QED and QCD corrections
discussed in Section~\ref{subsec:QED_QCD_corr}.
The measurements are sufficiently precise to require the presence of
quantum corrections associated with the 
virtual exchange of top quarks, gauge bosons and Higgses.

%%%%%%%%%%%%%%%%   FIGURE  %%%%%%%%%%%%%%%%%%%
\begin{figure}[thbp]
%\centerline{\framebox(50,20){FIGURE}}
\centerline{\mbox{\epsfysize=16.0cm\epsffile{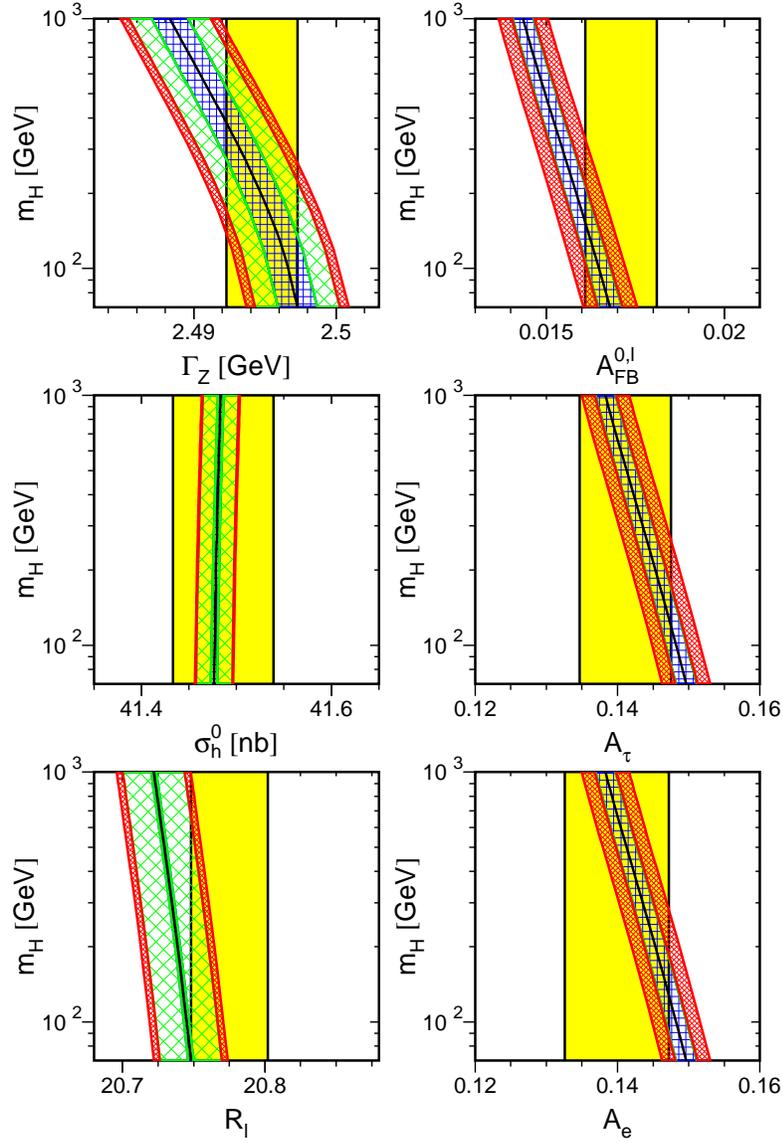}}}
%\vspace{-1.5cm}
\caption{Comparison \protect\cite{lepewwg:97}
of LEP measurements with the SM predictions as a
function of $m_H$. \hfill\hfil}
\label{fig:lep_results1}
\end{figure}

%%%%%%%%%%%%%%%%%%%%% END FIGURE %%%%%%%%%%%%%%%%%%%%%%%%%%

%%%%%%%%%%%%%%%%   FIGURE  %%%%%%%%%%%%%%%%%%%
\begin{figure}[htbp]
%\centerline{\framebox(50,20){FIGURE}}
\centerline{\mbox{\epsfysize=16.0cm\epsffile{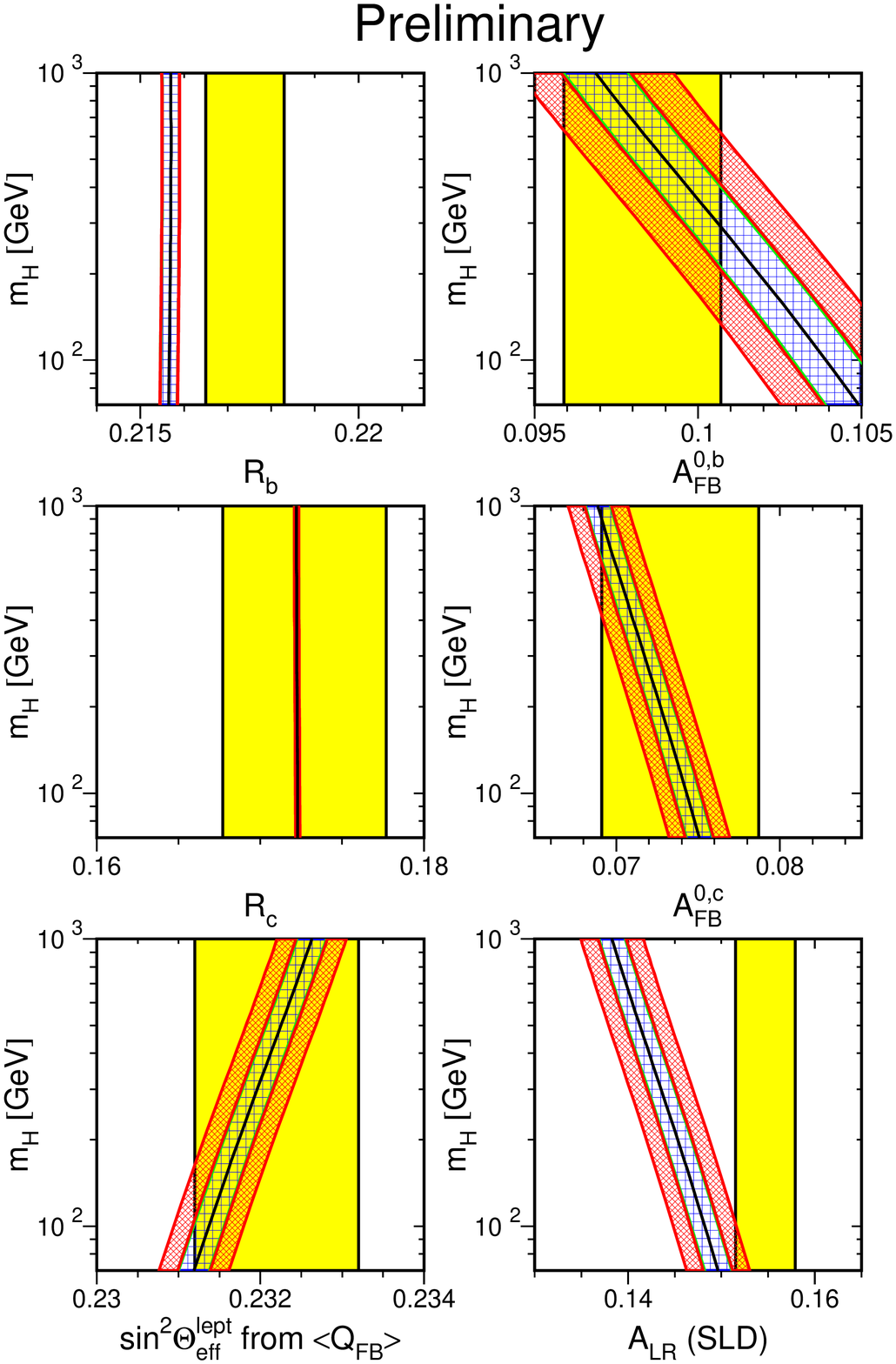}}}
%\vspace{-1.35cm}
\caption{Comparison \protect\cite{lepewwg:97}
of LEP measurements with the SM predictions as a
function of $m_H$.
Also shown is the SLD measurement of \protect{$\cA^0_{\mbox{\rms LR}}$}.
\hfill\hfil}
\label{fig:lep_results2}
\end{figure}
%%%%%%%%%%%%%%%%%%%%% END FIGURE %%%%%%%%%%%%%%%%%%%%%%%%%%

Figures \ref{fig:lep_results1} and \ref{fig:lep_results2}, 
taken from Ref.~\citen{lepewwg:97}, 
compare different
LEP measurements with the corresponding SM predictions
as a function of the Higgs mass;
also shown is the SLD measurement of $\cA^0_{\mbox{\rms LR}}$.
The cross-hatch pattern parallel to the axes indicates the variation of 
the SM prediction with $m_t = 175.6\pm 5.5$ GeV;
the coarse diagonal cross-hatch pattern corresponds to a variation of 
the strong coupling in the range
$\alpha_s(M_Z^2) = 0.118\pm 0.003$; and the dense diagonal
cross-hatching to the variation of $\alpha^{-1}(M_Z^2) = 128.896\pm 0.090$.
The experimental errors on the measured parameters 
are indicated as vertical bands.
For the comparison of $R_b$ with the SM the value of $R_c$ has been fixed
to the SM prediction.
The overall agreement is good.
As can be seen, the asymmetries are the most sensitive observables
to $m_H$.

Notice that the uncertainty on $\alpha(M_Z^2)^{-1}$ introduces
a severe limitation on the accuracy of the SM predictions.
To improve the present determination of $\alpha(M_Z^2)^{-1}$
one needs to perform a good measurement of
$\sigma(e^+e^-\to \mbox{\rm hadrons})$, as a function of the centre--of--mass
energy, in the whole kinematical range spanned by DA$\Phi$NE, a 
tau--charm factory and the B factories.
This would result in a much stronger constraint on the Higgs mass.

%%%%%%%%%%%%%%% Table %%%%%%%%%%%%%%%%%
\begin{table}[bht]
\begin{center}
\caption{Results from the global electroweak fits \protect\cite{lepewwg:97}
to LEP data alone,
to all data except the direct measurements of $m_t$ and $M_W$ at
Tevatron and LEP2, and to all data. \hfill\hfil
\label{tab:EW_fit}}
\vspace{0.2cm}
\begin{tabular}{|c|c|c|c|}
\hline
& LEP only & All data except & All data
\\
& ($M_W$ included) & $m_t$ and $M_W$ &  
\\ \hline %\hline
$m_t$ \,\, (GeV) & $158 {\,}^{+14}_{-11}$ &
$157{\,}^{+10}_{-9}$ & $173.1\pm 5.4$
\\
$m_H$ \,\, (GeV) & $83 {\,}^{+168}_{-49}$ &
$41{\,}^{+64}_{-21}$ & $115{\,}^{+116}_{-66}$
\\
$\log{(m_H)}$ & $1.92 {\,}^{+0.48}_{-0.39}$ &
$1.62{\,}^{+0.41}_{-0.31}$ & $2.06{\,}^{+0.30}_{-0.37}$
\\
$\alpha_s(M_Z^2)$ & $0.121\pm 0.003$ & $0.120\pm 0.003$ & 
$0.120\pm 0.003$
\\ \hline
$\chi^2/\mbox{\rm d.o.f.}$ & $8/9$ & $14/12$ & $17/15$
\\ \hline %\hline

$\sin^2{\theta^{\mbox{\rms lept}}_{\mbox{\rms eff}}}$ &
$0.23188\pm 0.00026$ & $0.23153\pm 0.00023$ & $0.23152\pm 0.00022$ 
\\
$1-M_W^2/M_Z^2$ & $0.2246\pm 0.0008$ & $0.2240\pm 0.0008$ &
$0.2231\pm 0.0006$ 
\\
$M_W$ \,\, (GeV) & $80.298\pm 0.043$ & $80.329\pm 0.041$ &
$80.375\pm 0.030$ 
\\ \hline
\end{tabular}
\end{center}
\end{table}
%%%%%%%%%%%% End Table %%%%%%%%%%%%%%%%

Table~\ref{tab:EW_fit} shows the constraints obtained on
$m_t$, $m_H$ and $\alpha_s(M_Z^2)$, from a global fit to the
electroweak data \cite{lepewwg:97}.
As the sensitivity to the Higgs mass is logarithmic, the fitted values of
$\log{(m_H)}$ are also quoted.
The bottom part of the table lists derived results for
$\sin^2{\theta^{\mbox{\rms lept}}_{\mbox{\rms eff}}}$,
$1 - M_W^2/M_Z^2$ and $M_W$.

Three different fits are shown. The first one uses only LEP data,
including the LEP2 determination of $M_W$.
The fitted value of the top mass is in good agreement with the
Tevatron measurement \cite{lepewwg:97}, $m_t= 175.6\pm 5.5$ GeV,
although slightly lower.
The data seems to prefer also a light Higgs.
There is a large correlation (0.76) between the fitted values of $m_t$
and $m_H$; the correlation would be much larger if the $R_b$ measurement
was not used ($R_b$ is insensitive to $m_H$).
The extracted value of the strong coupling agrees very well
with the world average value \cite{PDG:96} $\alpha_s(M_Z^2) = 0.118\pm 0.003$.

The second fit includes all electroweak data except the direct measurements
of $m_t$ and $M_W$, performed at Tevatron and LEP2. The fitted values for these
two masses agree well with the direct determinations. The indirect measurements
clearly prefer low $m_t$ and low $m_H$.

%%%%%%%%%%%%%%% Table %%%%%%%%%%%%%%%%%
\begin{table}[tbh]
\begin{center}
\caption{Summary \protect\cite{lepewwg:97}
of measurements included in the global analysis of electroweak
data. The third and fourth column show the fitted values obtained within the SM
and the associated pulls (difference between measurement and fit in units of
the measurement error). \hfill\hfil
\label{tab:pulls}}
\vspace{0.2cm}
\begin{tabular}{|c|c|c|c|}
\hline
& Measurement & SM fit & Pull
\\ \hline  %\hline
%\underline{LEP} &&&\\
$M_Z$ \,\, (GeV) & $91.1867\pm 0.0020$ & $91.1866$ & $0.0$
\\ 
$\Gamma_Z$ \,\, (GeV) & $2.4948\pm 0.0025$ & $2.4966$ & $-0.7$
\\ 
$\sigma^0_{\mbox{\rms had}}$ \,\, (nb) & $41.486\pm 0.053$ & $41.467$ & $0.4$
\\ 
$R_l$ & $20.775\pm 0.027$ & $20.756$ & $0.7$
\\ 
$\cA^{0,l}_{\mbox{\rms FB}}$ & $0.0171\pm 0.0010$ & $0.0162$ & $0.9$
\\ 
$\cP_\tau$ & $-0.1411\pm 0.0064$ & $-0.1470$ & $0.9$
\\ 
${4\over 3}\cA^{0,\tau}_{\mbox{\rms FB,Pol}}$ & $-0.1399\pm 0.0073$ & $-0.1470$ & 
$1.0$
\\
$R_b$ & $0.2170\pm 0.0009$ & $0.2158$ & $1.3$
%   $0.2174\pm 0.0009$ & $0.2158$ & $1.8$
\\
$R_c$ & $0.1734\pm 0.0048$ & $0.1723$ & $0.2$
%   $0.1727\pm 0.0050$ & $0.1723$ & $-0.1$
\\
$\cA^{0,b}_{\mbox{\rms FB}}$ & $0.0984\pm 0.0024$ & $0.1031$ & $-2.0$
\\ 
$\cA^{0,c}_{\mbox{\rms FB}}$ & $0.0741\pm 0.0048$ & $0.0736$ & $0.1$
\\
$\cP_b$ & $-0.900\pm 0.050$ & $-0.935$ & $0.7$
\\ 
$\cP_c$ & $-0.650\pm 0.058$ & $-0.668$ & $0.3$
\\ 
$\sin^2{\theta^{\mbox{\rms lept}}_{\mbox{\rms eff}}}$ \,\, 
($\langle Q_{\mbox{\rms FB}}\rangle$) &
$0.2322\pm 0.0010$ & $0.23152$ & $0.7$
\\ 
$\sin^2{\theta^{\mbox{\rms lept}}_{\mbox{\rms eff}}}$ \,\, 
($\cA^{0}_{\mbox{\rms LR}}$) &
$0.23055\pm 0.00041$ & $0.23152$ & $-2.4$
\\ 
$M_W$ \,\, (GeV) & $80.43\pm 0.08$ & $80.375$ & $0.7$
% $M_W$ \,\, (GeV) & $80.48\pm 0.14$ & $80.375$ & $0.8$
\\
%\\ \hline
%\underline{SLD} &&&\\
% $\sin^2{\theta^{\mbox{\rms lept}}_{\mbox{\rms eff}}}$ \,\, 
% ($\cA^{0}_{\mbox{\rms LR}}$) &
% $0.23055\pm 0.00041$ & $0.23152$ & $-2.4$\\ 
% $R_b$ & $0.2124\pm 0.0029$ & $0.2158$ & $-1.1$\\
% $R_c$ & $0.1810\pm 0.0145$ & $0.1723$ & $0.6$\\
% $\cP_b$ & $-0.900\pm 0.050$ & $-0.935$ & $0.7$\\ 
% $\cP_c$ & $-0.650\pm 0.058$ & $-0.668$ & $0.3$\\ 
% \\ \hline
% \underline{$p\bar p$} &&&\\ 
% $M_W$ \,\, (GeV) & $80.41\pm 0.09$ & $80.375$ & $0.4$\\ 
$m_t$ \,\, (GeV) & $175.6\pm 5.5$ & $173.1$ & $0.4$
% \\ \hline
% \underline{$\nu N$} &&&
\\ 
$1 - M_W^2/M_Z^2$ \,\, ($\nu N$) & $0.2254\pm 0.0037$ & $0.2231$ & $0.6$
\\ \hline
\end{tabular}
\end{center}
\end{table}
%%%%%%%%%%%% End Table %%%%%%%%%%%%%%%%

The best constraints on $m_H$ are obtained in the last fit, which includes all
available data. Taking into account additional theoretical uncertainties
(not included in Table~\ref{tab:EW_fit})
due to the missing higher--order corrections,
the global fit results in the upper bound \cite{lepewwg:97}:
\bel{eq:M_H}
m_H < 420 \;\mbox{\rm GeV} \qquad (95\% \, \mbox{\rm CL}) \, .
\ee

The quality of the global electroweak fit can better appreciated
in Table~\ref{tab:pulls}, where the input measurements are compared with
the resulting SM values. The pulls indicate the differences between the
measurements and the corresponding fit values in units of the experimental
errors. The global agreement is quite good.
The distribution of the pulls, with a few $2\sigma$ deviations, is more or less
what one should expect statistically.
As already mentioned before, the largest discrepancies occur in 
$\cA^{0}_{\mbox{\rms LR}}$ and $\cA^{0,b}_{\mbox{\rms FB}}$.

\section{Summary}
\label{sec:summary}

The SM provides a beautiful theoretical framework which is able to
accommodate all our present knowledge on electroweak interactions.
It is able to explain any single experimental fact  and, in some cases,
it has successfully passed very precise
tests at the 0.1\% to 1\% level. 
However, there are still  pieces of the SM Lagrangian which so far
have not been experimentally analyzed in any precise way.

The gauge self-couplings are presently being investigated at LEP2, through the
study of the $e^+e^-\to W^+W^-$ production cross-section.
The $V-A$ ($\nu_e$-exchange in the $t$ channel) contribution generates
an unphysical growing of the  cross-section with the centre-of-mass energy, 
which is compensated through a delicate gauge cancellation with the
$e^+e^-\to\gamma, Z\to W^+W^-$ amplitudes.
The recent LEP2 measurements of $\sigma(e^+e^-\to W^+W^-)$ at 172 GeV,
in good agreement with the SM,
have provided already convincing evidence \cite{lepewwg:97,miquel}
for the contribution coming from the $ZWW$ vertex.
With more statistics, it will be possible to make a detailed investigation
of the gauge--boson self-interactions.

The study of this process will also provide a more accurate measurement of
$M_W$, allowing to improve the precision of the neutral--current analyses.
The present LEP2 determination,
$M_W = 80.48\pm 0.14$ GeV, is already competitive with the value
$M_W = 80.41\pm 0.09$ GeV obtained in $p\bar p$ colliders,
but its error is still too large compared with the 41 MeV sensitivity
achieved  with the indirect SM fit of electroweak data.
To achieve this sort of accuracy is one of the main goals of LEP2.

The Higgs particle is the main missing block of the SM framework.
The data provide a clear confirmation of the assumed pattern of
spontaneous symmetry breaking, but do not prove the minimal Higgs
mechanism embedded in the SM.
At present, a relatively light Higgs seems to be preferred by the
indirect precision tests.
LEP2 first and later LHC will try to find out whether such scalar field
exists.

In spite of its enormous phenomenological success, the SM leaves too many
unanswered questions to be considered as a complete description of the
fundamental forces.
We do not understand yet why fermions are replicated in three
(and only three)
nearly identical copies? Why the pattern of masses and mixings
is what it is?  Are the masses the only difference among the three
families? What is the origin of the SM flavour structure?
Which dynamics is responsible for the observed CP violation?

Clearly, we need more experiments in order to learn what kind of
physics exists beyond the present SM frontiers.
We have, fortunately, a very promising and exciting future
ahead of us.

\section*{Acknowledgements}

I would like to thank the organizers for this enjoyable meeting,
and Arcadi Santamar\'{\i}a for his help with the figures.
I am indebted to the LEP Electroweak Working Group and the SLD Heavy 
Flavour Group for their work, which I have extensively used to write
these lectures.
This work has been supported in part by CICYT (Spain) under grant
No. AEN-96-1718.

%%%%%%%%%%%%%%%%%%%%%%%%%%%%  REFERENCES %%%%%%%%%%%%%%%%%%%%

\end{document}